\documentclass[aps,eqsecnum,preprint,floats,epsf,epsfig,nofootinbib]{revtex4}
\usepackage{epsfig}

\begin{document}
\def\be{\begin{eqnarray}}
\def\en{\end{eqnarray}}
\def\non{\nonumber}
\def\la{\langle}
\def\ra{\rangle}
\def\A{{\cal A}}
\def\B{{\cal B}}
\def\c{{\cal C}}
\def\d{{\cal D}}
\def\e{{\cal E}}
\def\p{{\cal P}}
\def\t{{\cal T}}
\def\nc{N_c^{\rm eff}}
\def\CP{{\it CP}~}
\def\CPP{{\it CP}}
\def\vp{\varepsilon}
\def\drho{\bar\rho}
\def\deta{\bar\eta}
\def\vma{{_{V-A}}}
\def\vpa{{_{V+A}}}
\def\J{{J/\psi}}
\def\ov{\overline}
\def\Lqcd{{\Lambda_{\rm QCD}}}
\def\pr{{ Phys. Rev.}~}
\def\prl{{ Phys. Rev. Lett.}~}
\def\pl{{ Phys. Lett.}~}
\def\np{{ Nucl. Phys.}~}
\def\zp{{ Z. Phys.}~}
\def\lsim{ {\ \lower-1.2pt\vbox{\hbox{\rlap{$<$}\lower5pt\vbox{\hbox{$\sim$}
}}}\ } }
\def\gsim{ {\ \lower-1.2pt\vbox{\hbox{\rlap{$>$}\lower5pt\vbox{\hbox{$\sim$}
}}}\ } }

\font\el=cmbx10 scaled \magstep2{\obeylines \hfill November, 2013}

\vskip 1.5 cm

\centerline{\large\bf Branching Fractions and Direct {\it CP} Violation}
\centerline{\large\bf in Charmless Three-body Decays of $B$ Mesons}
\bigskip
\centerline{\bf Hai-Yang Cheng$^{1}$, Chun-Khiang Chua$^{2}$}
\medskip
\centerline{$^1$ Institute of Physics, Academia Sinica}
\centerline{Taipei, Taiwan 115, Republic of China}
\medskip
\centerline{$^2$ Department of Physics and Center for High Energy Physics}
\centerline{Chung Yuan Christian University}
\centerline{Chung-Li, Taiwan 320, Republic of China}

\medskip

\bigskip
\bigskip
\bigskip
\centerline{\bf Abstract}
\bigskip

\small

Charmless three-body decays of $B$ mesons are studied using a simple model based on the framework of the factorization approach. Hadronic three-body decays receive both resonant and nonresonant contributions.
Dominant nonresonant contributions to tree-dominated three-body decays arise
from the $b\to u$ tree transition which can be evaluated using heavy meson chiral perturbation theory valid in the soft meson limit.
For penguin-dominated decays, nonresonant signals come mainly from the penguin amplitude governed by the matrix elements of scalar densities $\la M_1M_2|\bar q_1 q_2|0\ra$.  We use the measurements of $\ov B^0\to K_SK_SK_S$ to constrain the nonresonant component of
$\la K\ov K|\bar ss|0\ra$. The
intermediate vector meson contributions to three-body decays are
identified through the vector current, while the scalar meson
resonances are mainly associated with the scalar density.
While the calculated direct \CP violation in $B^-\to K^+K^-K^-$ and $B^-\to \pi^+\pi^-\pi^-$ decays agrees well with experiment in both magnitude and sign, the predicted \CP asymmetries in $B^-\to \pi^- K^+K^-$ and $B^-\to K^-\pi^+\pi^-$ have incorrect signs when confronted with experiment. It has been conjectured recently that a possible resolution to this \CP puzzle may rely on final-state rescattering of $\pi^+\pi^-$ and $K^+K^-$. Assuming a large strong phase associated with the matrix element $\la K\pi|\bar sq|0\ra$ arising from some sort of power corrections, we fit it to the data of $K^-\pi^+\pi^-$ and find a correct sign for $\pi^- K^+K^-$.  We predict some testable \CP violation in $\ov B^0\to K^+K^-\pi^0$ and  $K^+K^-K_S$.
In the low mass regions of the Dalitz plot, we find that the regional \CP violation is indeed largely enhanced with respect to the inclusive one, though it is still significantly below the data. In this work, strong phases arise from effective Wilson coefficients, propagators of resonances and the matrix element of scalar density $\la M_1M_2|\bar q_1q_2|0\ra$.

\pagebreak

\section{Introduction}

Recently, LHCb has measured direct \CP violation in charmless three-body decays of $B$ mesons \cite{LHCb:Kppippim,LHCb:pippippim,deMiranda} and found evidence of \CP asymmetries in $B^+\to\pi^+\pi^+\pi^-$ (4.9$\sigma$), $B^+\to K^+K^+K^-$ (3.7$\sigma$) and $B^+\to K^+K^-\pi^+$ (3.2$\sigma$) and a 2.8$\sigma$ signal of \CP violation in $B^+\to K^+\pi^+\pi^-$ (see Table \ref{tab:CPdata}). Direct \CP violation in two-body resonances in the Dalitz plot has been seen at $B$ factories. For example, both BaBar \cite{BaBar:Kmpippim} and Belle \cite{Belle:Kmpippim} have claimed evidence of partial rate asymmetries in the channel $B^\pm\to \rho^0(770)K^\pm$ in the Dalitz-plot analysis of $B^\pm\to K^\pm\pi^\mp\pi^\pm$.
The inclusive \CP asymmetry in three-body decays results from the interference of the two-body resonances and three-body nonresonant decays and through the tree-penguin interference. \CP asymmetries in certain local regions of the phase space are likely to be greater than the integrated inclusive ones. Indeed, LHCb has also observed large asymmetries in localized regions of phase space \cite{LHCb:Kppippim,LHCb:pippippim,deMiranda}. For example,
\be \label{eq:KKKlocalCP}
\A_{CP}^{\rm region}(K^+K^-K^-)=-0.226\pm0.020\pm0.004\pm0.007
\en
for $m^2_{K^+K^- \rm ~high}<15$ GeV$^2$ and $1.2<m^2_{K^+K^- \rm ~low}<2.0$ GeV$^2$,
\be \label{eq:KpipilocalCP}
\A_{CP}^{\rm region}(K^-\pi^+\pi^-)=0.678\pm0.078\pm0.032\pm0.007
\en
for $m^2_{K^-\pi^+ \rm ~high}<15$ GeV$^2$ and $0.08<m^2_{\pi^+\pi^- \rm ~low}<0.66$ GeV$^2$,
\be \label{eq:KKpilocalCP}
\A_{CP}^{\rm region}(K^+K^-\pi^-)=-0.648\pm0.070\pm0.013\pm0.007
\en
for $m^2_{K^+K^-}<1.5$ GeV$^2$, and
\be \label{eq:pipipilocalCP}
\A_{CP}^{\rm region}(\pi^+\pi^-\pi^-)=0.584\pm0.082\pm0.027\pm0.007
\en
for $m^2_{\pi^-\pi^- \rm ~low}<0.4$ GeV$^2$ and $m^2_{\pi^+\pi^- \rm ~high}>15$ GeV$^2$.
Hence, significant signatures of \CP violation were found in the above-mentioned low mass regions devoid of most of the known resonances.

Three-body decays of heavy mesons are more complicated than the
two-body case as they receive both resonant and nonresonant
contributions.  The
analysis of these decays using the Dalitz plot technique enables
one to study the properties of various vector and scalar resonances.
Indeed, most of the quasi-two-$B$ decays are
extracted from the Dalitz-plot analysis of three-body ones. Three-body hadronic $B$ decays involving a vector meson or a charmed meson in the final state also have been observed at $B$ factories. In this work we shall focus on  charmless $B$ decays into three pseudoscalar mesons.

It is known that the nonresonant signal in charm decays is small,
less than 10\% \cite{PDG}. In the past years, many of the
charmless $B$ to three-body decay modes have been measured at $B$
factories and studied using the Dalitz-plot analysis. The
measured fractions and the corresponding branching fractions of
nonresonant components for some of three-body $B$ decay modes are
summarized in Table \ref{tab:BRexpt}.
We see that the nonresonant fraction is about $\sim (70-100)\%$
in $B\to KKK$ decays, $\sim (17- 40)\%$  in $B\to K\pi\pi$ decays, and $\sim$ 35\% in the $B\to\pi\pi\pi$
decay. Hence, the nonresonant three-body decays play an essential
role in penguin-dominated $B$ decays. While this is striking in view of the rather
small nonresonant background in three-body charm decays, it is not
entirely unexpected because the energy release scale in weak $B$
decays is of order 5 GeV, whereas the major resonances lie in the
energy region of 0.77 to 1.6 GeV. Consequently, it is likely that
three-body $B$ decays may receive sizable nonresonant contributions.
It is important to
understand and identify the underlying mechanism for nonresonant
decays.

\begin{table}[t]
\caption{Experimental results of direct \CP asymmetries (in \%) for various charmless
three-body $B$ decays \cite{HFAG,LHCb:Kppippim,LHCb:pippippim}. }
\begin{ruledtabular} \label{tab:CPdata}
\begin{tabular}{l  ccc c }
 Final state~~ & BaBar & Belle & LHCb & Average \\ \hline
 $K^+K^-K^-$ & $-1.7^{+1.9}_{-1.4}\pm1.4$ &  &  $-4.3\pm0.9\pm0.3\pm0.7$ & $-3.7\pm1.0$
 \\
 $(K^+K^-K^-)_{\rm NR}$ & $6.0\pm4.4\pm1.3$ &  &  $$ & $6.0\pm4.8$
 \\
 $K^-K_SK_S$ & $4^{+4}_{-5}\pm2$ & & & $4^{+4}_{-5}$ \\
 $K^+K^-\pi^-$ & $0\pm10\pm3$ & & $-14.1\pm4.0\pm1.8\pm0.7$ & $-11.9\pm4.1$  \\
 $K^-\pi^+\pi^-$ & $2.8\pm2.0\pm2.0\pm1.2$ & $4.9\pm2.6\pm2.0$ & $3.2\pm0.8\pm0.4\pm0.7$ & $3.3\pm1.0$ \\
 $K^-\pi^+\pi^0$ & $-3.0^{+4.5}_{-5.1}\pm5.5$ & $7\pm11\pm1$ & & $0.0^{+5.9}_{-6.1}$ \\
 $(K^-\pi^+\pi^0)_{\rm NR}$ & $10\pm16\pm8$ & $$ & & $10\pm18$ \\
 $K^-\pi^0\pi^0$ & $-6\pm6\pm4$ & & & $-6\pm7$ \\
 $\ov K^0\pi^+\pi^-$ & $-1\pm5\pm1$& & & $-1\pm5$ \\
 $\pi^+\pi^-\pi^-$ &   $3.2\pm4.4\pm3.1^{+2.5}_{-2.0}$ & & $11.7\pm2.1\pm0.9\pm0.7$ &  $10.5\pm2.2$ \\
 $(\pi^+\pi^-\pi^-)_{\rm NR}$ &   $-14\pm14^{+18}_{-~8}$ & & $$ &  $-14^{+23}_{-16}$ \\
\end{tabular}
\end{ruledtabular}
\end{table}

\begin{table}[t]
\caption{Branching fractions of various charmless three-body decays
of $B$ mesons. The fractions and the corresponding branching
fractions of nonresonant (NR) components are included whenever
available. The first, second and third entries are BaBar, Belle and LHCb
results, respectively. } \label{tab:BRexpt}
 \footnotesize{
\begin{ruledtabular}
\begin{tabular}{l c c c c c}
 Decay & $\B$($10^{-6}$) &  $\B$$_{_{\rm NR}}(10^{-6})$  & NR fraction(\%) & Resonances &  Ref. \\
\hline
 $B^- \to\pi^+\pi^-\pi^-$ &  $15.2\pm0.6\pm1.3$ &
$5.3\pm0.7^{+1.3}_{-0.8}$  & $34.9\pm4.2^{+8.0}_{-4.5}$ & $\rho^0,\rho^0(1450)$ & \cite{BaBarpipipi} \\
 & -- & -- & -- & $f_0(1370),f_2(1270)$ \\
 $B^-\to K^-\pi^+\pi^-$ & $54.4\pm1.1\pm4.6$ &
 $9.3\pm1.0^{+6.9}_{-1.7}$ \footnotemark[1] & $17.1\pm1.7^{+12.4}_{-~1.8}$ & $K^{*0},K_0^{*0},\rho^0,\omega$ & \cite{BaBar:Kmpippim} \\
 & $48.8\pm1.1\pm3.6$ & $16.9\pm1.3^{+1.7}_{-1.6}$ & $34.0\pm2.2^{+2.1}_{-1.8}$ & $f_0(980),K_2^{*0},f_2(1270)$ & \cite{Belle:Kmpippim} \\
 $B^-\to K^-\pi^0\pi^0$ & $16.2\pm1.2\pm1.5$ & & & $K^{*-},f_0(980)$ & \cite{BaBarKmpi0pi0} \\
 & -- & -- & -- \\
 $B^-\to K^+K^-\pi^-$ & $5.0\pm0.5\pm0.5$ & & & & \cite{BaBarKpKmpim} \\
 & $<13$  &  & & & \cite{Belle2004} \\
 $B^-\to K^+K^-K^-$ & $33.4\pm0.5\pm0.9$ \footnotemark[2] &
 $22.8\pm2.7\pm7.6$  & $68.3\pm8.1\pm22.8$ & $\phi,f_0(980),f_0(1500)$ & \cite{BaBarKKK} \\
 & $30.6\pm1.2\pm2.3$ \footnotemark[2] & $24.0\pm1.5\pm1.5$
 & $78.4\pm5.8\pm7.7$ & $f_0(1710),f'_2(1525)$ &   \cite{BelleKpKpKm} \\
 $B^-\to K^-K_SK_S$ & $10.1\pm0.5\pm0.3$ & $19.8\pm3.7\pm2.5$ & $\sim$196 & $f_0(980),f_0(1500)$  & \cite{BaBarKKK}  \\
 & $13.4\pm1.9\pm1.5$ & & & $f_0(1710),f'_2(1525)$ & \cite{Belle2004} \\
 \hline
 $ \ov B^0\to \ov K^0\pi^+\pi^-$ & $50.2\pm1.5\pm1.8$ & $11.1^{+2.5}_{-1.0}\pm0.9$ & $22.1^{+2.8}_{-2.0}\pm2.2$ & $f_0(980),\rho^0,K^{*+}$ &  \cite{BaBarK0pippim}\\
 & $47.5\pm2.4\pm3.7$ & $19.9\pm2.5^{+1.7}_{-2.0}$ & $41.9\pm5.1^{+1.5}_{-2.6}$ &  $K_0^{*+},f_2(1270)$ & \cite{BelleK0pipi} \\
 $\ov B^0\to K^-\pi^+\pi^0$ & $38.5\pm1.0\pm3.9$ & $7.6\pm0.5\pm1.0$ \footnotemark[3] & $19.7\pm1.4\pm3.3$ & $\rho^+,\rho^+(1450)$ & \cite{BaBarKppimpi0} \\
 & $36.6^{+4.2}_{-4.3}\pm3.0$ & $5.7^{+2.7+0.5}_{-2.5-0.4}<9.4$ & $<25.7$ & $K^{*(0,-)},K_0^{*(0,-)}$ & \cite{BelleKppimpi0} \\
 $\ov B^0\to \stackrel{(-)}{K^0} K^\mp\pi^\pm$ \footnotemark[4] & $6.4\pm1.0\pm0.6$ & & & &\cite{BaBarKpK0pim} \\
 & $<18$ & & & & \cite{Belle2004} \\
 & $6.4\pm0.9\pm0.4\pm0.3$ & & & & \cite{LHCb:KKpi} \\
 $\ov B^0\to K^+K^-\pi^0$ & -- & & & &  \\
 & $2.17\pm0.60\pm0.24$ & & &  & \cite{Belle:KpKmpi0}\\
 $\ov B^0\to K^+K^-\ov K^0$ & $25.4\pm0.9\pm0.8$ \footnotemark[2] & $33\pm5\pm9$  & $\sim$130 & $\phi,f_0(980),f_0(1500)$ &  \cite{BaBarKKK} \\
 & $28.3\pm3.3\pm4.0$ & & & $f_0(1710),f'_2(1525)$ & \cite{Belle2004} \\
 & $19.1\pm1.5\pm1.1\pm0.8$ & & & & \cite{LHCb:KKpi} \\
 $\ov B^0\to K_SK_SK_S$ & $6.19\pm0.48\pm0.19$ & $13.3^{+2.2}_{-2.3}\pm2.2$ & $\sim$215 & $f_0(980),f_0(1710)$ & \cite{BaBarKsKsKs} \\
 & $4.2^{+1.6}_{-1.3}\pm0.8$ & & & $f_2(2010)$ & \cite{Belle2004} \\
\end{tabular}
\end{ruledtabular} }
 \footnotetext[1]{The branching fraction for the phase-space nonresonant  is $(2.4\pm0.5^{+1.3}_{-1.5})\times 10^{-6}$.}
 \footnotetext[2]{Contributions from $\chi_{c0}$ are excluded.}
  \footnotetext[3]{The branching fraction for the phase-space nonresonant  is $(2.8\pm0.5\pm0.4)\times 10^{-6}$.}
   \footnotetext[4]{It is the sum of $\ov K^0K^+\pi^-$ and $K^0K^-\pi^+$.}
\end{table}

Consider the nonresonant contributions to the three-body $B$ decay
$B\to P_1P_2P_3$. Under the factorization hypothesis, one of the
nonresonant components arises from the transitions $B\to P_1
P_2$ with an emission of $P_3$.  The nonresonant background in charmless three-body $B$
decays due to the transition $B\to P_1 P_2$ has been studied
extensively
\cite{Deshpande,Fajfer1,Fajfer2,Deandrea1,Deandrea,Fajfer3} based
on heavy meson chiral perturbation theory (HMChPT)
\cite{Yan,Wise,Burdman}. However, the predicted rates of nonresonant decays due to $B\to P_1P_2$ transition alone already exceed the measured total branching fractions for the tree-dominated modes e.g. $\pi^-\pi^+\pi^-$ and $\pi^-K^+K^-$. For example, the branching fraction
of the nonresonant rate of $B^-\to \pi^+\pi^-\pi^-$ estimated using HMChPT is found to be of order $75\times 10^{-6}$, which is even larger than the total branching fraction of order $15\times 10^{-6}$ (see Table \ref{tab:BRexpt}).  The issue has to do with the applicability of HMChPT. When it is applied to three-body decays, two of the final-state pseudoscalars  have to be soft. If the soft meson result is assumed to be the same in the whole Dalitz plot, the decay rate will be
greatly overestimated. To overcome this issue, we have proposed in \cite{CCS:nonres} to parameterize the momentum dependence of nonresonant amplitudes induced by $b\to u$ transition in an exponential form so that the HMChPT results are recovered in the soft pseudoscalar meson limit.

However, the nonresonant background in $B\to P_1P_2$ transition
does not suffice to account for the experimental observation that nonresonant contributions dominate in the penguin-dominated decays $B\to KKK$ and $B\to K\pi\pi$. As we have emphasized in \cite{CCS:nonres}, this implies that the nonresonant amplitude is also penguin dominated and governed by the matrix elements, e.g., $\la K\ov K|\bar ss|0\ra$ and $\la K\pi|\bar sq|0\ra$. That is, the matrix element of scalar density should have a large nonresonant component. In \cite{CCS:nonres} we have used the $\ov B^0\to
K_SK_SK_S$ mode in conjunction with the  mass spectrum in $\ov
B^0\to K^+K^-\ov K^0$ to fix the nonresonant contribution to $\la
K\ov K|\bar ss|0\ra$.

Besides the nonresonant background, it is necessary to study resonant
contributions to three-body decays. The
intermediate vector meson contributions to three-body decays are
identified through the vector current, while the scalar meson
resonances are mainly associated with the scalar density. They can also contribute to
the three-body matrix element $\la P_1P_2|J_\mu|B\ra$.
Resonant effects are conventionally described in terms of the usual Breit-Wigner formalism. In this manner we are able to identify the relevant
resonances which contribute to the three-body decays of interest and
compute the rates of $B\to VP$ and $B\to SP$. In conjunction with
the nonresonant contribution, we are ready to calculate the total
rates for three-body decays.

There are several competing approaches for describing charmless hadronic two-body decays of $B$ mesons, such as  QCD factorization (QCDF) \cite{BBNS},  pQCD \cite{Li} and soft-collinear effective theory \cite{SCET}. Measurements of \CP asymmetries will allow us to discriminate between different models and  improve the approach. For example, in the heavy quark limit, the predicted \CP asymmetries for the penguin-dominated modes $\ov B^0\to K^-\pi^+$, $K^{*-}\pi^+$, $K^-\rho^+$, $\ov B_s^0\to K^+\pi^-$ have incorrect signs when confronted with experiment \cite{CC:BCP,CC:Bud}. In the approach of QCDF, their signs can be flipped into the right direction by considering $1/m_b$  power corrections from penguin annihilation. Therefore, even an information on the sign of \CP asymmetries will be very valuable.

The recent LHCb measurements of inclusive and local direct \CP asymmetries in charmless $B\to P_1P_2P_3$ decays \cite{LHCb:Kppippim,LHCb:pippippim,deMiranda} provide a new testground of the factorization approach. Let's first check the signs of \CP violation. The observed negative relative sign of \CP asymmetries between $B^-\to \pi^-\pi^+\pi^-$ and $B^-\to K^-K^+K^-$ and between $B^-\to K^-\pi^+\pi^-$ and $B^-\to \pi^-K^+K^-$ is in accordance with what expected from U-spin symmetry which enables us to relate the $\Delta S=0$ amplitude to the $\Delta S=1$ one. However, symmetry arguments alone do not tell us the relative sign of \CP asymmetries between $\pi^-\pi^+\pi^-$ and $\pi^-K^+K^-$ and between $K^-\pi^+\pi^-$ and  $K^-K^+K^-$.  Based on a realistic model calculation we find positive relative signs which are in contradiction to the LHCb experiment. How to resolve this \CP enigma becomes a very important issue in the study of hadronic 3-body decays.
The LHCb observation of the correlation of the \CP violation between the decays, $\A_{CP}(\pi^-\pi^+\pi^-)\approx -\A_{CP}(\pi^-K^+K^-)$ and $\A_{CP}(K^-\pi^+\pi^-)\approx -\A_{CP}(K^-K^+K^-)$, has led to the conjecture that $\pi^+\pi^-\leftrightarrow K^+K^-$
rescattering may play an important role in the generation of the strong phase
difference needed for such a violation to occur.

In this work we shall follow the framework of \cite{CCS:nonres} to update the analysis of three-body decays and explore inclusive and regional \CP violation in detail. We take
the factorization approximation as a working hypothesis rather
than a first-principles starting point as factorization has not been proved for three-body $B$ decays. Therefore, we
shall work in the phenomenological factorization model rather than
in the established theories such as  QCDF, pQCD or
soft-collinear effective theory. \footnote{For the study of $B\to PPP$ decays in different approaches, the reader is referred to \cite{ElBennich:2009da,London}.}
For \CP violation, we will focus on direct \CP asymmetry and will not discuss mixing-induced \CP violation in, for example, $B^0\to K^+K^-K_S$ and
$K_SK_SK_S$. This topic  has been discussed in \cite{CCS:nonres,CCS:KKKS}.

The layout of the present paper is as follows. We shall first discuss
the decay $B\to\pi\pi\pi$ in Sec. II in order to fix the parameter for describing the nonresonant background at the tree level. We discuss this mode in detail to set up the framework for studying resonant and nonresonant  contributions. Then in Sec. III we proceed to $B\to KKK$ decays to emphasize the importance of nonresonant penguin contributions to penguin-dominated modes. The three-body channels $B\to K\pi\pi$ and $B\to KK\pi$ are discussed in Secs. IV and V, respectively. In Sec. VI, we determine the
rates for $B\to VP$ and $B\to SP$ and compare our results with the
approach of  QCD factorization. Inclusive and localized \CP asymmetries are addressed in Sec. VII. Sec. VIII contains our conclusions. Some of the input parameters used in this work are collected in Appendix A.
Factorizable amplitudes for some of $B\to PPP$ decays not discussed previously in \cite{CCS:nonres} are shown in Appendix B.

\section{$B\to \pi\pi\pi$ decays}

For three-body $B$ decays, the $b\to sq\bar q$ penguin transitions
contribute to the final states with odd number of kaons, namely,
$KKK$ and $K\pi\pi$, while $b\to uq\bar q$ tree and $b\to dq\bar
q$ penguin transitions contribute to final states with even number
of kaons, e.g. $KK\pi$ and $\pi\pi\pi$. We shall discuss
the decay $B\to\pi\pi\pi$ first in order to fix the parameter needed for describing the nonresonant background at the tree level and then $B\to KKK$ to fix the unknown parameter for the nonresonant penguin contribution. Finally we proceed to discuss $B\to K\pi\pi$ and $B\to KK\pi$ channels.

Under the factorization
hypothesis, the decay amplitudes are given by
 \be \label{eq:factamp}
 \la P_1P_2P_3|{\cal H}_{\rm eff}|B\ra
 =\frac{G_F}{\sqrt2}\sum_{p=u,c}\lambda_p^{(r)} \la P_1P_2P_3|T_p^{(r)}|B\ra,
 \en
where $\lambda_p^{(r)}\equiv V_{pb} V^*_{pr}$ with $r=d,s$. For
$KKK$ and $K\pi\pi$ modes, $r=s$ and for $KK\pi$ and $\pi\pi\pi$
channels, $r=d$. The Hamiltonian $T_p^{(r)}$ has the expression
\cite{BBNS}
 \be \label{eq:Tp}
 T_p^{(r)}&=&
 a_1 \delta_{pu} (\bar u b)_{V-A}\otimes(\bar r u)_{V-A}
 +a_2 \delta_{pu} (\bar r b)_{V-A}\otimes(\bar u u)_{V-A}
 +a_3(\bar r b)_{V-A}\otimes\sum_q(\bar q q)_{V-A}
 \non\\
 &&+a^p_4\sum_q(\bar q b)_{V-A}\otimes(\bar r q)_{V-A}
   +a_5(\bar r b)_{V-A}\otimes\sum_q(\bar q q)_{V+A}
       \non\\
 &&-2 a^p_6\sum_q(\bar q b)_{S-P}\otimes(\bar r q)_{S+P}
 +a_7(\bar r b)_{V-A}\otimes\sum_q\frac{3}{2} e_q (\bar q q)_{V+A}
 \non\\
 &&-2a^p_8\sum_q(\bar q b)_{S-P}\otimes\frac{3}{2} e_q
             (\bar r q)_{S+P}
 +a_9(\bar r b)_{V-A}\otimes\sum_q\frac{3}{2}e_q (\bar q q)_{V-A}\non\\
 &&+a^p_{10}\sum_q(\bar q b)_{V-A}\otimes\frac{3}{2}e_q(\bar r
 q)_{V-A},
 \en
with $(\bar q q')_{V\pm A}\equiv \bar q\gamma_\mu(1\pm\gamma_5)
q'$, $(\bar q q')_{S\pm P}\equiv\bar q(1\pm\gamma_5) q'$ and a
summation over $q=u,d,s$ being implied. For the effective Wilson
coefficients, we shall follow \cite{CCS:nonres} to use
 \be \label{eq:ai}
 && a_1\approx0.99\pm0.037 i,\quad a_2\approx 0.19-0.11i, \quad a_3\approx -0.002+0.004i, \quad a_5\approx
 0.0054-0.005i,  \non \\
 && a_4^u\approx -0.03-0.02i, \quad a_4^c\approx
 -0.04-0.008i,\quad
 a_6^u\approx -0.06-0.02i, \quad a_6^c\approx -0.06-0.006i,
 \non\\
 &&a_7\approx 0.54\times 10^{-4} i,\quad a_8^u\approx (4.5-0.5i)\times
 10^{-4},\quad
 a_8^c\approx (4.4-0.3i)\times
 10^{-4},   \\
 && a_9\approx -0.010-0.0002i,\quad
 a_{10}^u \approx (-58.3+ 86.1 i)\times10^{-5},\quad
 a_{10}^c \approx (-60.3 + 88.8 i)\times10^{-5}, \non
 \en
for typical $a_i$ at the renormalization scale $\mu=m_b/2=2.1$~GeV. The strong phases of the effective Wilson coefficients arise from vertex corrections and penguin contractions calculated in the QCD factorization approach \cite{BBNS}.

\subsection{$B^-\to\pi^+\pi^-\pi^-$ decay}

The factorizable tree-dominated $B^-\to\pi^+\pi^-\pi^-$ decay reads
\be \label{eq:A3pi}
 \la \pi^+ \pi^- \pi^-|T_p|B^-\ra &=&
 \la \pi^+ \pi^-|(\bar u b)_{V-A}|B^-\ra \la \pi^-|(\bar d u)_{V-A}|0\ra
 \left[a_1 \delta_{pu}+a^p_4+a_{10}^p-(a^p_6+a^p_8) r_\chi^\pi\right]
 \non\\
&& + \la \pi^-|(\bar db)_{V-A}|B^-\ra \la\pi^+\pi^-|(\bar
uu)_{V-A}|0\ra\Big[a_2\delta_{pu}+a_3+a_5+a_7+a_9\Big] \non \\
&& + \la \pi^-|(\bar db)_{V-A}|B^-\ra \la\pi^+\pi^-|(\bar
dd)_{V-A}|0\ra\Big[a_3+a_4^p+a_5-{1\over 2}(a_7+a_9+a_{10}^p)\Big] \non \\
&& + \la \pi^-|(\bar db)_{V-A}|B^-\ra \la\pi^+\pi^-|(\bar
ss)_{V-A}|0\ra\Big[a_3+a_5-{1\over 2}(a_7+a_9)\Big] \non \\
&& +\la \pi^-|\bar db|B^-\ra \la\pi^+\pi^-|\bar dd|0\ra
(-2a_6^p+a_8^p) \non \\
&& +\la \pi^-\pi^+\pi^-|(\bar du)_{V-A}|0\ra \la0|(\bar
ub)_{V-A}|B^-\ra(a_1\delta_{pu}+a_4^p+a_{10}^p) \non \\
&& +\la \pi^-\pi^+\pi^-|\bar d(1+\gamma_5)u|0\ra \la0|\bar
u\gamma_5b|B^-\ra(2a_6^p+2a_8^p),
\en
where $r_\chi^\pi(\mu)=2{m_\pi^2\over m_b(\mu)(m_d(\mu)-m_u(\mu))}$.
Since there are two identical $\pi^-$ mesons in this decay, one
should take into account the identical particle effects. For
example,
 \be
\la \pi^+\pi^-|(\bar u b)_{V-A}|B^-\ra \la \pi^-|(\bar d u)_{V-A}|0\ra
 &=& \la \pi^+(p_1)\pi^-(p_2)|(\bar u b)_{V-A}|B^-\ra \la \pi^-(p_3)|(\bar d u)_{V-A}|0\ra
 \non \\
 &+& \la \pi^+(p_1)\pi^-(p_3)|(\bar u b)_{V-A}|B^-\ra \la \pi^-(p_2)|(\bar d
 u)_{V-A}|0\ra,
 \non \\
 \en
and a factor of ${1\over 2}$ should be put in the decay rate. Note that $\la \pi^+\pi^-|(\bar dd)_{V-A}|0\ra=-\la \pi^+\pi^-|(\bar uu)_{V-A}|0\ra$ due to isospin symmetry. The matrix element $\la \pi^+\pi^-|(\bar ss)_{V-A}|0\ra$ is suppressed by the OZI rule.

Under the factorization approach, the $B^-\to
\pi^+\pi^-\pi^-$ decay amplitude consists of three distinct
factorizable terms: (i) the current-induced process with a meson
emission, $\la B^-\to \pi^+\pi^-\ra\times \la 0\to \pi^-\ra$,
(ii) the transition process,  $\la B^-\to \pi^-\ra\times \la
0\to \pi^+\pi^-\ra$, and (iii) the annihilation process $\la
B^-\to 0\ra\times \la 0\to \pi^+\pi^-\pi^-\ra$, where $\la A\to
B\ra$ denotes a $A\to B$ transition matrix element.
We shall consider the nonresonant background and resonant contributions separately.

\subsubsection{Nonresonant background}

For the current-induced process, the three-body matrix
element $\la \pi^+ \pi^-|(\bar u b)_{V-A}|B^-\ra$ has
the general expression~\cite{LLW}
\be \label{eq:romegah}
 \la \pi^+ (p_1) \pi^-(p_2)|(\bar u b)_{V-A}|B^-\ra
 &=&i r
 (p_B-p_1-p_2)_\mu+i\omega_+(p_2+p_1)_\mu+i\omega_-(p_2-p_1)_\mu
 \non\\
 &&+h\,\epsilon_{\mu\nu\alpha\beta}p_B^\nu (p_2+p_1)^\alpha
 (p_2-p_1)^\beta.
\en
The form factors $r$,
$\omega_\pm$ and $h$ can be evaluated in the framework of heavy meson chiral perturbation theory (HMChPT) \cite{LLW}.
However, this will lead to decay rates that are too large, in disagreement
with experiment \cite{Cheng:2002qu}. The heavy meson chiral
Lagrangian given in \cite{Yan,Wise,Burdman} is needed to compute
the strong $B^*BP$, $B^*B^*P$ and $BBPP$ vertices. The results for
the form factors read \cite{LLW,Fajfer1}
 \be \label{eq:r&omega}
 \omega_+ &=& -{g\over f_\pi^2}\,{f_{B^*}m_{B^*}\sqrt{m_Bm_{B^*}}\over
 s_{23}-m_{B^*}^2}\left[1-{(p_B-p_1)\cdot p_1\over
 m_{B^*}^2}\right]+{f_B\over 2f_\pi^2}, \non \\
 \omega_- &=& {g\over f_\pi^2}\,{f_{B^*}m_{B^*}\sqrt{m_Bm_{B^*}}\over
 s_{23}-m_{B^*}^2}\left[1+{(p_B-p_1)\cdot p_1\over
 m_{B^*}^2}\right], \non \\
 r &=& {f_B\over 2f_\pi^2}-{f_B\over
 f_\pi^2}\,{p_B\cdot(p_2-p_1)\over
 (p_B-p_1-p_2)^2-m_B^2}+{2gf_{B^*}\over f_\pi^2}\sqrt{m_B\over
 m_{B^*}}\,{(p_B-p_1)\cdot p_1\over s_{23}-m_{B^*}^2} \non \\
 &-& {4g^2f_B\over f_\pi^2}\,{m_Bm_{B^*}\over
 (p_B-p_1-p_2)^2-m_B^2}\,{p_1\!\cdot\!p_2-p_1\!\cdot\!(p_B-p_1)\,p_2\!\cdot\!
 (p_B-p_1)/m_{B^*}^2 \over s_{23}-m_{B^*}^2 },
 \en
where $s_{ij}\equiv (p_i+p_j)^2$, $f_\pi=132$ MeV, $g$ is a heavy-flavor independent strong
coupling which can be extracted from the CLEO measurement of the
$D^{*+}$ decay width, $|g|=0.59\pm0.01\pm0.07$ \cite{CLEOg}. We
shall follow \cite{Yan} to fix its sign to be negative. It follows that
\be \label{eq:AHMChPT}
 A_{\rm current-ind}^{\rm HMChPT} &\equiv&\la \pi^-(p_3)|(\bar s
 u)_{V-A}|0\ra \la   \pi^+ (p_1) \pi^-(p_2)|(\bar u b)_{V-A}|B^-\ra \non\\
 &=& -\frac{f_\pi}{2}\left[2 m_3^2 r+(m_B^2-s_{12}-m_3^2) \omega_+
 +(s_{23}-s_{13}-m_2^2+m_1^2) \omega_-\right].
\en

However, as pointed out before, the predicted nonresonant rates based on HMChPT are unexpectedly too large for tree-dominated decays. For example, the branching fraction of nonresonant $B^-\to \pi^+\pi^-\pi^-$ is found to be of order $75\times 10^{-6}$, which is one order of magnitude larger than the BaBar result of $\sim 5.3\times 10^{-6}$ (see Table \ref{tab:BRexpt}). The issue has to do with the applicability of HMChPT. In order to
apply this approach, two of the final-state pseudoscalars in $B\to
P_1P_2$ transition have to be soft. The momentum of the soft
pseudoscalar should be smaller than the chiral symmetry breaking
scale of order 1 GeV. For three-body charmless $B$ decays, the
available phase space where chiral perturbation theory is
applicable is only a small fraction of the whole Dalitz plot.
Therefore, it is not justified to apply chiral and heavy quark
symmetries to a certain kinematic region and then generalize it to
the region beyond its validity.  If the soft meson result is
assumed to be the same in the whole Dalitz plot, the decay rate
will be greatly overestimated.
Following \cite{CCS:nonres}, we shall assume the momentum dependence of nonresonant amplitudes in an exponential
form, namely,
\be \label{eq:ADalitz}
  A_{\rm current-ind}=A_{\rm current-ind}^{\rm
  HMChPT}\,e^{-\alpha_{_{\rm NR}}
p_B\cdot(p_1+p_2)}e^{i\phi_{12}},
\en
so that the HMChPT results are recovered in the soft meson limit
$p_1,~p_2\to 0$. That is, the nonresonant amplitude in the soft
meson region is described by HMChPT, but its energy dependence
beyond the chiral limit is governed by the exponential term
$e^{-\alpha_{_{\rm NR}} p_B\cdot(p_1+p_2)}$.  In what follows, we
shall use the tree-dominated $B^-\to\pi^+\pi^-\pi^-$ decay data to
fix the unknown parameter $\alpha_{_{\rm NR}}$. Besides the nonresonant contribution from the current-induced process, the matrix elements $\la \pi^+\pi^-|\bar q\gamma_\mu q|0\ra$ and $\la \pi^+\pi^-|\bar d d|0\ra$ also receive nonresonant contributions. In principle, the weak vector form factor of the former matrix element can be related to the charged pion electromagnetic (e.m.) form factors.
However, unlike the kaon case which will be discussed below, the time-like e.m. form factors of the pions are not well measured enough allowing us to determine the nonresonant parts. Therefore, we shall only consider the resonant contribution to $\la \pi^+\pi^-|\bar q\gamma_\mu q|0\ra$. As for the matrix element $\la \pi^+\pi^-|\bar d d|0\ra$, it can be related to $\la K^+K^-|\bar ss|0\ra$ to be discussed below via SU(3) flavor symmetry. Nevertheless, it is suppressed by the smallness of the penguin Wilson coefficients $a_6$ and $a_8$. Therefore, the nonresonant component of $B^-\to\pi^-\pi^+\pi^-$ is predominated by the current-induced process, and its measurement provides an ideal place to constrain the parameter $\alpha_{_{\rm NR}}$, which turns out to be
\be \label{eq:alphaNR}
 \alpha_{_{\rm NR}}=0.081^{+0.015}_{-0.009}\,{\rm GeV}^{-2}.
\en
This is very close to the naive expectation of $\alpha_{_{\rm
NR}}\sim {\cal O}(1/(2m_B\Lambda_\chi))$ based on the dimensional
argument. The phase $\phi_{12}$ of the nonresonant amplitude in
the $(\pi^+\pi^-)$ system will be set to zero for simplicity.

\subsubsection{Resonant contributions}

In general, vector meson and scalar resonances contribute to the two-body matrix elements $\langle P_1P_2|V_\mu|0\rangle$ and $\langle
P_1P_2|S|0\rangle$, respectively. \footnote{The two-body matrix element $\la P_1P_2|V_\mu|0\ra$ sometimes can also receive contributions from scalar resonances. For example, both $K^*$ and $K_0^*(1430)$ contribute to the matrix element $\la K^-\pi^+|(\bar sd)_{V-A}|0\ra$, see Eq. (\ref{eq:m.e.pole}).}
They can also contribute to the
three-body matrix element $\langle P_1P_2|J_\mu|B\rangle$.
Resonant effects are described in terms of the usual Breit-Wigner
formalism. More precisely,
\begin{eqnarray}
  \la \pi^+(p_1)\pi^-(p_2)|(\bar ub)_\vma|B^-\ra^R &=& \sum_i\langle \pi^+\pi^-|V_i\rangle
 {1\over s-m_{V_i}^2+im_{V_i}\Gamma_{V_i}}\langle V_i|(\bar ub)_\vma|B^-\rangle \non \\
&+& \sum_i\langle \pi^+\pi^-|S_i\rangle
 {-1\over s-m_{S_i}^2+im_{S_i}\Gamma_{S_i}}\langle S_i|(\bar ub)_\vma|B^-\rangle, \non \\
  \langle \pi^+\pi^-|\bar q\gamma_\mu q|0\rangle^R &=& \sum_i\langle \pi^+\pi^-|V_i\rangle
 {1\over s-m_{V_i}^2+im_{V_i}\Gamma_{V_i}}\langle V_i|\bar q\gamma_\mu q|0\rangle, \nonumber \\
 \langle \pi^+\pi^-|\bar dd|0\rangle^R &=& \sum_i\langle \pi^+\pi^-|S_i\rangle
 {-1\over s-m_{S_i}^2+im_{S_i}\Gamma_{S_i}}\langle S_i|\bar dd|0\rangle,
 \end{eqnarray}
where $V_i=\phi,\rho,\omega,\cdots$ and
$S_i=f_0(980),f_0(1370),f_0(1500),\cdots$.
It follows that
\begin{eqnarray} \label{eq:m.e.pole}
  \la \pi^+(p_1)\pi^-(p_2)|(\bar ub)_\vma|B^-\ra^R &=& \sum_i
{g^{V_i\to \pi^+\pi^-}\over
s_{12}-m_{V_i}^2+im_{V_i}\Gamma_{V_i}}\sum_{\rm
pol}\vp^*\cdot
(p_1-p_2)\la V_i|(\bar ub)_\vma|B^-\ra \non \\
&-& \sum_i{g^{{S_i}\to \pi^+\pi^-}\over
s_{12}- m_{S_i}^2+im_{S_i}\Gamma_{S_i}}\la
S_i|(\bar ub)_\vma|B^-\ra, \non \\
  \langle \pi^+(p_1)\pi^-(p_2)|\bar q\gamma_\mu q|0\rangle^R &=& \sum_i
 {g^{V_i\to \pi^+\pi^-}\over s-m_{V_i}^2+im_{V_i}\Gamma_{V_i}}\sum_{\rm
pol}\vp^*\cdot(p_1-p_2)\langle V_i|\bar
q\gamma_\mu q|0\rangle, \nonumber \\
 \langle \pi^+\pi^-|\bar dd|0\rangle^R &=& -\sum_i
 {g^{S_i\to \pi^+\pi^-}\over s- m_{S_i}^2+im_{S_i}\Gamma_{S_i}}\langle S_i|\bar dd|0\rangle.
 \end{eqnarray}

Using the decay constants defined by
\be \label{eq:decayc}
\la S|\bar q_2q_1|0\ra=m_S\bar f_S, \quad \la P(p)|\bar q_2\gamma_\mu\gamma_5 q_1|0\ra=-if_Pp_\mu, \quad
 \la V(p)|\bar q_2\gamma_\mu q_1|0\ra=f_Vm_V\vp_\mu^*,
\en
and form factors defined by \footnote{We follow \cite{BSW} for the  $B\to P$ and $B\to V$ transition form factors. Form factors for $B\to S$ transitions are defined in \cite{CCH}.}
\be \label{eq:FF}
 \la P(p')|V_\mu|B(p)\ra &=& \left((p+p')_\mu-{m_B^2-m_P^2\over q^2}\,q_ \mu\right)
F_1^{BP}(q^2)+{m_B^2-m_P^2\over q^2}q_\mu\,F_0^{BP}(q^2), \non \\
\la S(p')|A_\mu|B(p)\ra &=&
-i\Bigg[\left((p+p')_\mu-{m_B^2-m_S^2\over q^2}\,q_ \mu\right)
F_1^{BS}(q^2)   +{m_B^2-m_S^2\over
q^2}q_\mu\,F_0^{BS}(q^2)\Bigg], \non \\
\la V(p',\vp)|V_\mu|B(p)\ra &=& {2\over
m_B+m_V}\,\epsilon_{\mu\nu\alpha \beta}\vp^{*\nu}p^\alpha p'^\beta
V(q^2),   \non \\
\la V(p',\vp)|A_\mu|B(p)\ra &=& i\Big[ (m_B+m_V)\vp^*_\mu
A_1^{BV}(q^2)-{\vp^*\cdot p\over m_B+m_V}\,(p+p')_\mu
A_2^{BV}(q^2) \non  \\ && -2m_V\,{\vp^*\cdot p\over
q^2}\,q_\mu\big[A_3^{BV}(q^2)-A_0^{BV}(q^2)\big]\Big],
 \en
where $P_\mu=(p+p')_\mu$, $q_\mu=(p-p')_\mu$, $A^{BV}_3(0)=A^{BV}_0(0)$, and
\be
A^{BV}_3(q^2)=\,{m_B+m_V\over 2m_V}\,A^{BV}_1(q^2)-{m_B-m_V\over
2m_V}\,A^{BV}_2(q^2),
\en
we are led to
 \be
 && \la \pi^+(p_1)\pi^-(p_2)|(\bar ub)\vma|B^-\ra^R ~\la \pi^-(p_3)|(\bar
du)_\vma|0\ra \non \\
&=& -\sum _i{f_\pi\over 2\sqrt{2}}\,{g^{\rho_i\to\pi^+\pi^-}\over
s_{12}-m_{\rho_i^2}+im_{\rho_i}\Gamma_{\rho_i}}(s_{13}-s_{23})\Big[
(m_B+m_{\rho_i})A_1^{B\rho_i}(q^2) \non \\
&&- {A_2^{B\rho_i}(q^2)\over m_B+m_{\rho_i}}
(m_B^2-s_{12})-2m_{\rho_i}[A_3^{B\rho_i}(q^2)-A_0^{B\rho_i}(q^2)]\Big] \non\\
&&- \sum _if_\pi{g^{{f_0}_i\to\pi^+\pi^-}\over
s_{12}-m_{{f_0}_i}^2+im_{{f_0}_i}\Gamma_{{f_0}_i}}(m_B^2-s_{12})F_0^{Bf_0^u}(q^2),
\en
with $q^2=(p_B-p_1-p_2)^2=p_3^2$, and
\be \label{eq:R2body}
 \la \pi^+(p_1)\pi^-(p_2)|\bar u\gamma_\mu u|0\ra^R
 &=&  -{1\over\sqrt{2}}\sum_i{m_{\rho_i}f_{\rho_i}g^{\rho_i\to \pi^+\pi^-}\over
 s_{12}-m_{\rho_i}^2+im_{\rho_i}\Gamma_{\rho_i}}\,(p_1-p_2)_\mu, \non \\
 \la \pi^+(p_1) \pi^-(p_2)|\bar d d|0\ra^R
 &=& -\sum_{i}\frac{m_{{f_0}_i} \bar f^d_{{f_0}_i} g^{{f_0}_i\to \pi^+\pi^-}}{s_{12}-m_{{f_0}_i}^2+i m_{{f_0}_i}\Gamma_{{f_0}_i}},
\en
the
scalar decay constant $\bar f_{{f_0}_i}^q$ is defined by $\la
{f_0}_i|\bar q q|0\ra=m_{{f_0}_i} \bar f^q_{{f_0}_i}$,
$g^{{f_0}_i\to \pi^+\pi^-}$ is the ${f_0}_i\to
\pi^+\pi^-$ strong coupling.
Hence, the relevant transition amplitudes are
\be \label{}
&& \la \pi^+(p_1)\pi^-(p_2)|(\bar uu)_{V-A}|0\ra^R
\la \pi^-(p_3)|(\bar d b)_{V-A}|B^-\ra
= -F_1^{B\pi}(s_{12})F^{\pi^+\pi^-}_R(s_{12})\left(s_{13}-s_{23}\right), \non \\
&& \la \pi^+(p_1)\pi^-(p_2)|\bar dd|0\ra^R \la \pi^-|\bar db|B^-\ra= - {m_B^2-m_\pi^2\over m_b-m_d}F_0^{B\pi}(s_{12})\sum_{i}\frac{m_{{f_0}_i} \bar f^d_{{f_0}_i} g^{{f_0}_i\to \pi^+\pi^-}}{s_{12}-m_{{f_0}_i}^2+i
 m_{{f_0}_i}\Gamma_{{f_0}_i}},
\en
with
\be
F^{\pi^+\pi^-}_R(s)={1\over\sqrt{2}}\sum_i{m_{\rho_i}f_{\rho_i}g^{\rho_i\to \pi^+\pi^-}\over
s- m_{\rho_i}^2+im_{\rho_i}\Gamma_{\rho_i}}.
\en

\subsubsection{Numerical results}

\begin{table}[t]
\caption{Branching fractions (in units of $10^{-6}$) of resonant and
nonresonant (NR) contributions to $B^-\to
\pi^+\pi^-\pi^-$. The nonresonant background is used as an input to
fix the parameter $\alpha_{_{\rm NR}}$ defined in Eq.
(\ref{eq:ADalitz}). Theoretical errors correspond to the uncertainties in (i)
$\alpha_{_{\rm NR}}$, (ii) $F^{B\pi}_0$, $\sigma_{_{\rm
NR}}$ and $m_s(\mu)=(90\pm 20) $MeV at $\mu=2.1$ GeV,  and (iii) $\gamma=(69.7^{+1.3}_{-2.8})^\circ$.  Experimental results are taken
from Table \ref{tab:BRexpt}.}
\begin{ruledtabular} \label{tab:pippimpim}
\begin{tabular}{l l l }
Decay mode~~
  &  BaBar \cite{BaBarpipipi}
  & Theory
  \\
  \hline
$\rho^0\pi^-$
  & $8.1\pm0.7\pm1.2^{+0.4}_{-1.1}$
  &  $6.7^{+0.0+0.4+0.1}_{-0.0-0.4-0.1}$ %
  \\
$\rho^0(1450)\pi^-$
  & $1.4\pm0.4\pm0.4^{+0.3}_{-0.7}$
  \\
$f_0(1370)\pi^-$
  & $2.9\pm0.5\pm0.5^{+0.7}_{-0.5}$
  & $1.6^{+0.0+0.0+0.0}_{-0.0-0.0-0.0}$ 
  \\
$f_0(980)\pi^-$
  & $<1.5$
  & $0.2^{+0.0+0.0+0.0}_{-0.0-0.0-0.0}$%
  \\
NR
  & $5.3\pm0.7\pm0.6^{+1.1}_{-0.5}$
  & input
  \\
\hline
Total
  & $15.2\pm0.6\pm1.2^{+0.4}_{-0.3}$
  & $16.1^{+1.9+1.0+0.2}_{-2.3-0.8-0.2}$%
  \\
\end{tabular}
\end{ruledtabular}
\end{table}

The strong coupling constants such as $g^{\rho(770)\to\pi^+\pi^-}$ and
$g^{f_0(980)\to\pi^+\pi^-}$ are determined from the measured
partial widths through the relations
 \be
 \Gamma_{S\to P_1P_2}={p_c\over 8\pi m_S^2}g_{S\to P_1P_2}^2,\qquad
 \Gamma_{V\to P_1P_2}={2\over 3}\,{p_c^3\over 4\pi m_V^2}g_{V\to P_1P_2}^2,
 \en
for scalar and vector mesons, respectively, where $p_c$ is the
c.m. momentum. The numerical results are
 \be \label{eq:g}
 && g^{\rho(770)\to\pi^+\pi^-}=6.0, \qquad\qquad\quad
\quad g^{K^*(892)\to K^+\pi^-}=4.59,\non \\
&& g^{f_0(980)\to\pi^+\pi^-}=1.33^{+0.29}_{-0.26}\,{\rm GeV},
\quad g^{K_0^*(1430)\to K^+\pi^-}=3.84\,{\rm GeV}.
 \en
Note that the neutral $\rho$ meson cannot decay into $\pi^0\pi^0$ owing to isospin invariance.
In determining the coupling of $f_0\to\pi^+\pi^-$, we have used
the partial width
 \be
 \Gamma(f_0(980)\to\pi^+\pi^-)=(34.2^{+13.9+8.8}_{-11.8-2.5})\,{\rm
 MeV}
 \en
measured by Belle \cite{Bellef0}. In this work, we shall specifically use $g^{f_0(980)\to\pi^+\pi^-}=1.18\,{\rm GeV}$ to have a better description of $B\to f_0(980)K$ channels in $B\to K\pi\pi$ decays.

The calculated branching fractions of resonant and nonresonant
contributions to $B^-\to \pi^+\pi^-\pi^-$ are summarized in
Table \ref{tab:pippimpim}. The theoretical errors shown there are
from the uncertainties in (i) the parameter $\alpha_{_{\rm NR}}$ [see Eq. (\ref{eq:alphaNR})]
which governs the momentum dependence of the nonresonant
amplitude, (ii) the strange quark mass $m_s$ for decay modes involving kaon(s), the form factor
$F^{B\pi}_0$ and the nonresonant parameter $\sigma_{_{\rm NR}}$ to be introduced below in  Eq. (\ref{eq:KKssme}), and
(iii) the unitarity angle $\gamma$.

We see from Table \ref{tab:pippimpim} that the decay $B^-\to\pi^+\pi^-\pi^-$ is dominated by the $\rho^0$ pole and the nonresonant contribution.
The calculated total branching fraction $(16.1^{+1.9}_{-2.3})\times 10^{-6}$ agrees well with experiment.

\subsection{$\ov B^0\to \pi^+\pi^-\pi^0$ decay}

\begin{table}[t]
\caption{Predicted branching fractions (in units of $10^{-6}$) of resonant and
nonresonant (NR) contributions to  $\ov
B^0\to \pi^+\pi^-\pi^0$.}
\begin{ruledtabular} \label{tab:pippimpi0}
\begin{tabular}{l l | l l }
Decay mode~~~~~~~~~~~~~
  & Theory~~~~~~~~~~~~~~~~~~~~~
& Decay mode
  & Theory
  \\
\hline
$\rho^+\pi^-$
  & $3.8^{+0.0+0.4+0.0}_{-0.0-0.3-0.0}$ %
& $\rho^0\pi^0$
  & $1.0^{+0.0+0.2+0.0}_{-0.0-0.1-0.0}$%
  \\
$\rho^-\pi^+$
  & 
  $13.8^{+0.0+3.5+0.1}_{-0.0-3.1-0.1}$
&$f_0(980)\pi^0$
  & 
  $0.004^{+0.000+0.001+0.000}_{-0.000-0.001-0.000}$
  \\
$\rho^\pm\pi^\mp$
  & 
  $17.8^{+0.0+3.6+0.1}_{-0.0-3.1-0.1}$
&  NR
  & $1.6^{+0.5+0.0+0.0}_{-0.6-0.0-0.0}$%
  \\
\hline
Total
  & $20.1^{+0.3+3.7+0.1}_{-0.3-3.3-0.1}$ %
&
  &
\end{tabular}
\end{ruledtabular}
\end{table}

The factorizable amplitude of $\ov B^0\to\pi^+\pi^-\pi^0$ is given by
\be \label{eq:pippimpi0}
 \la \pi^0 \pi^+ \pi^-|T_p|\ov B^0\ra &=&
 \la \pi^+ \pi^0|(\bar u b)_{V-A}|\ov B^0\ra \la \pi^-|(\bar d u)_{V-A}|0\ra
 \left[a_1 \delta_{pu}+a^p_4+a_{10}^p-(a^p_6+a^p_8) r_\chi^\pi\right]
 \non\\
  &+&  \la \pi^+ \pi^-|(\bar d b)_{V-A}|\ov B^0\ra \la \pi^0|(\bar u u)_{V-A}|0\ra
 \Big[a_2 \delta_{pu}-a_4^p+(a_6^p-{1\over 2}a_8^p)r_\chi^\pi \non \\
 &+& {3\over 2}(a_7+a_9)+{1\over 2}a_{10}^p\Big]
 +
 \la \pi^+ |(\bar u b)_{V-A}|\ov B^0\ra \la \pi^-\pi^0|(\bar d u)_{V-A}|0\ra
 \left[a_1 \delta_{pu}+a^p_4+a_{10}^p\right]
 \non\\
&+&  \la \pi^0|(\bar db)_{V-A}|\ov B^0\ra \la\pi^+\pi^-|(\bar
uu)_{V-A}|0\ra\Big[a_2\delta_{pu}-a_4^p+{3\over 2}(a_7+a_9)+{1\over 2}a_{10}^p\Big] \non \\
&+& \la \pi^0|\bar db|\ov B^0\ra \la\pi^+\pi^-|\bar dd|0\ra
(-2a_6^p+a_8^p).
\en
It is obvious that while $B^-\to
\pi^+\pi^-\pi^-$ is dominated by the $\rho^0$ resonance, the decay
$\ov B^0\to \pi^+\pi^-\pi^0$ receives intermediate $\rho^\pm$ and $\rho^0$ pole contributions. As a consequence, the $\pi^+\pi^-\pi^0$ mode has a rate larger than $\pi^+\pi^-\pi^-$ even though the former does not have two identical particles in the final state and moreover it involves a $\pi^0$ meson. Note that the calculated branching fractions of $\ov B^0\to \rho^\pm\pi^\mp,\rho^0\pi^0$ shown in Table \ref{tab:pippimpi0} are consistent with the data (in units of $10^{-6}$), $23.0\pm2.3$ and $2.0\pm0.5$, respectively, measured from other processes \cite{HFAG}. The nonresonant rate in $\ov B^0\to\pi^+\pi^-\pi^0$  is fairly small because it is expected to be about four times smaller than that in $B^-\to\pi^+\pi^-\pi^-$. This is confirmed by a realistic calculation.

In Sec. V.C we shall explore the possibility if the large rate of $\ov B^0\to K^+K^-\pi^0$ observed by Belle recently \cite{Belle:KpKmpi0} can arise from the decay $\ov B^0\to\pi^+\pi^-\pi^0$ followed by final-state rescattering of $\pi^+\pi^-\to K^+K^-$.

\section{$B\to KKK$ decays}

\subsection{$B^-\to K^+K^-K^-$ decay}
The factorizable penguin-dominated $B^-\to K^+K^-K^-$
decay amplitude is given by
\be  \label{eq:AKpKmKm}
 \la K^+ K^- K^-|T_p|B^-\ra&=&
 \la K^+K^-|(\bar u b)_{V-A}|B^-\ra \la K^-|(\bar s u)_{V-A}|0\ra
 \left[a_1 \delta_{pu}+a^p_4+a_{10}^p-(a^p_6+a^p_8) r_\chi^K\right]
 \non\\
 &&+\la K^-|(\bar s b)_{V-A}| B^-\ra
                   \la K^+ K^-|(\bar u u)_{V-A}|0\ra
    (a_2\delta_{pu}+a_3+a_5+a_7+a_9)
                   \non\\
 &&+\la K^-|(\bar s b)_{V-A}| B^-\ra
                   \la K^+ K^-|(\bar d d)_{V-A}|0\ra
    \bigg[a_3+a_5-\frac{1}{2}(a_7+a_9)\bigg]
    \non\\
 &&+ \la K^-|(\bar s b)_{V-A}| B^-\ra
                   \la K^+ K^-|(\bar s s)_{V-A}|0\ra
    \bigg[a_3+a^p_4+a_5-\frac{1}{2}(a_7+a_9+a^p_{10})\bigg]
    \non\\
 &&+\la K^-|\bar s b|B^-\ra
       \la K^+ K^-|\bar s s|0\ra
       (-2 a^p_6+a^p_8)
       \non\\
  &&  +\la K^+ K^-K^-|(\bar s u)_{V-A}|0\ra
     \la 0|(\bar u b)_{V-A}|B^-\ra
       \bigg(a_1\delta_{pu}+a^p_4+ a^p_{10}\bigg)
       \non\\
 &&  + \la K^+ K^-K^-|\bar s(1+\gamma_5) u|0\ra
       \la 0|\bar u\gamma_5 b|B^-\ra
       (2a^p_6+2a^p_8).
 \en
For the current-induced process with a kaon emission, the form factors $r,~\omega_\pm$ and $h$ for the three-body matrix element $\la K^+K^-|(\bar u b)_{V-A}|B^-\ra$ [see Eq. (\ref{eq:romegah})] evaluated in the framework of HMChPT are the same as that of Eq. (\ref{eq:r&omega}) except that $B^*$ is replaced by $B^*_s$. As explained in the last section, the available phase space where chiral perturbation theory
is applicable is only a small fraction of the whole Dalitz plot. Therefore, we have proposed to parameterize the $b\to u$ transition-induced
nonresonant amplitude in an exponent form given in Eq. (\ref{eq:ADalitz}). The unknown parameter
$\alpha_{_{\rm NR}}$ is  determined from the data of the
tree-dominated decay $B^-\to\pi^+\pi^-\pi^-$ and is given by Eq. (\ref{eq:alphaNR}).

In addition to the $b\to u$ tree transition, we need to consider the
nonresonant contributions to the $b\to s$ penguin amplitude
\begin{eqnarray}
 A_1 &=& \langle K {}^-(p_1)|(\bar s b)_{V-A}|B {}^-\rangle
  \langle K^+(p_2) K^-(p_3)|(\bar qq)_{V-A}|0\rangle, \nonumber \\
 A_2 &=& \langle K {}^-(p_1)|\bar s b|B {}^-\rangle
       \langle K^+(p_2) K^-(p_3)|\bar s s|0\rangle.
\end{eqnarray}
The two-kaon creation matrix element can be expressed in terms of
time-like kaon current form factors as
 \be \label{eq:KKweakff}
 \la K^+(p_{K^+}) K^-(p_{K^-})|\bar q\gamma_\mu q|0\ra
 &=& (p_{K^+}-p_{K^-})_\mu F^{K^+K^-}_q,
 \non\\
 \la K^0(p_{K^0}) \ov K^0(p_{\bar K^0})|\bar q\gamma_\mu q|0\ra
 &=& (p_{K^0}-p_{\bar K^0})_\mu F^{K^0\bar K^0}_q.
 \en
The weak vector form factors $F^{K^+K^-}_q$ and $F^{K^0\bar
K^0}_q$ can be related to the kaon electromagnetic (e.m.) form
factors $F^{K^+K^-}_{\rm em}$ and $F^{K^0\bar K^0}_{\rm em}$ for the
charged and neutral kaons, respectively. Phenomenologically, the
e.m. form factors receive resonant and nonresonant contributions
and can be expressed by
 \be \label{eq:KKemff}
 F^{K^+K^-}_{\rm em}= F^{KK}_\rho+F^{KK}_\omega+F^{KK}_\phi+F_{NR}, \qquad
 F^{K^0\bar K^0}_{\rm em}= -F^{KK}_\rho+F^{KK}_\omega+F^{KK}_\phi+F_{NR}'.
 \en
It follows from Eqs. (\ref{eq:KKweakff}) and (\ref{eq:KKemff})
that
 \be
 F^{K^+K^-}_u&=&F^{K^0\bar K^0}_d=F^{KK}_\rho+3 F^{KK}_\omega+\frac{1}{3}(3F_{NR}-F'_{NR}),
 \non\\
 F^{K^+K^-}_d&=&F^{K^0\bar K^0}_u=-F^{KK}_\rho+3 F^{KK}_\omega,
 \non\\
 F^{K^+K^-}_s&=&F^{K^0\bar K^0}_s=-3 F^{KK}_\phi-\frac{1}{3}(3 F_{NR}+2F'_{NR}),
 \label{eq:FKKisospin}
 \en
where use of isospin symmetry has been made.

The resonant and nonresonant terms in Eq. (\ref{eq:KKemff}) can be
parameterized as
 \be
 F_{h}(s_{23})=\frac{c_h}{m^2_h-s_{23}-i m_h \Gamma_h},
 \qquad
 F^{(\prime)}_{NR}(s_{23})=\left(\frac{x^{(\prime)}_1}{s_{23}}
 +\frac{x^{(\prime)}_2}{s_{23}^2}\right)
 \left[\ln\left(\frac{s_{23}}{\tilde\Lambda^2}\right)\right]^{-1},
 \en
with $\tilde\Lambda\approx 0.3$ GeV. The expression for the
nonresonant form factor is motivated by the asymptotic constraint
from pQCD, namely, $F(t)\to (1/t)[\ln(t/\tilde \Lambda^2)]^{-1}$
in the large $t$ limit \cite{Brodsky}. The unknown parameters
$c_h$, $x_i$ and $x'_i$ are fitted from the kaon e.m. data, giving
the best fit values (in units of GeV$^2$ for $c_h$) ~\cite{DKK}:
\begin{equation}
\begin{array}{lll}
c_\rho=3c_\omega=c_\phi=0.363,
  & c_{\rho(1450)}=7.98\times 10^{-3},\ \
  & c_{\rho(1700)}=1.71\times10^{-3},\ \
\\
c_{\omega(1420)}=-7.64\times 10^{-2},
  & c_{\omega(1650)}=-0.116,
  & c_{\phi(1680)}=-2.0\times10^{-2},
\\
\end{array}
\label{eq:cj}
\end{equation}
and
\begin{eqnarray}
x_1=-3.26~{\rm GeV}^2, \qquad x_2=5.02~{\rm GeV}^4,
 \qquad x'_1=0.47~{\rm GeV}^2,
 \qquad x'_2=0.
\label{eq:xy}
\end{eqnarray}
Note that the form factors $F_{\rho,\omega,\phi}$ in
Eqs.~(\ref{eq:KKemff}) and (\ref{eq:FKKisospin}) include the
contributions from the vector mesons
$\rho(770),\,\rho(1450),\,\rho(1700)$,
$\omega(782),\,\omega(1420),\,\omega(1650),$ $\phi(1020)$ and
$\phi(1680)$.
As a cross check, following the derivation of the resonant component of $\la\pi^+\pi^-|\bar u\gamma_\mu u|0\ra$ in Eq. (\ref{eq:R2body}) we obtain the resonant contributions to the $K^+K^-$ transition form factors
\be
F^{K^+K^-}_{u,R}(s) &=& -{1\over\sqrt{2}}\left(\sum_i{m_{\rho_i}f_{\rho_i}g^{\rho_i\to K^+K^-}\over
 s-m_{\rho_i}^2+im_{\rho_i}\Gamma_{\rho_i}}+\sum_i{m_{\omega_i}f_{\omega_i}g^{\omega_i\to K^+K^-}\over
 s-m_{\omega_i}^2+im_{\omega_i}\Gamma_{\omega_i}}\right), \non \\
F^{K^+K^-}_{d,R}(s) &=& {1\over\sqrt{2}}\left(\sum_i{m_{\rho_i}f_{\rho_i}g^{\rho_i\to K^+K^-}\over
 s-m_{\rho_i}^2+im_{\rho_i}\Gamma_{\rho_i}}-\sum_i{m_{\omega_i}f_{\omega_i}g^{\omega_i\to K^+K^-}\over
 s-m_{\omega_i}^2+im_{\omega_i}\Gamma_{\omega_i}}\right), \non \\
F^{K^+K^-}_{s,R}(s) &=& -\sum_i{m_{\phi_i}f_{\phi_i}g^{\phi_i\to K^+K^-}\over
 s-m_{\phi_i}^2+im_{\phi_i}\Gamma_{\phi_i}}.
\en
Using the quark model result $g^{\rho\to K^+K^-}:g^{\omega\to K^+K^-}:g^{\phi\to K^+K^-}=1:1:-1/\sqrt{2}$ to fix the relative sign of strong couplings and noting that $g^{\phi\to K^+K^-}=-4.54$ determined from the measured $\phi\to K^+K^-$ rate, we find $c_\phi=-{1\over 3}m_\phi f_\phi g^{\phi\to K^+K^-}=0.340$ in agreement with $c_\phi=0.363$ obtained from a fit to the kaon e.m. data.

The use of the equation of motion thus leads to
 \be
 A_1 &=&  (s_{12}-s_{13}) F_1^{BK}(s_{23}) F^{K^+K^-}_q (s_{23}), \non\\
 A_2 &=& {m_B^2-m_K^2\over
 m_b-m_s}F_0^{BK}(s_{23})f_s^{K^+K^-}(s_{23}),
 \en
where the matrix element $f_s^{K^+K^-}$ receives both resonant
and non-resonant contributions:
\be \label{eq:KKssme}
 \la K^+(p_2) K^-(p_3)|\bar s s|0\ra
 &\equiv& f^{K^+K^-}_s(s_{23})=\sum_{i}\frac{m_{{f_0}_i} \bar f^s_{{f_0}_i} g^{{f_0}_i\to K^+K^-}}{m_{{f_0}_i}^2-s_{23}-i
 m_{{f_0}_i}\Gamma_{{f_0}_i}}+f_s^{NR},
 \non\\
 f_s^{NR}&=&\frac{v}{3}(3 F_{NR}+2F'_{NR})+\sigma_{_{\rm NR}}
 e^{-\alpha s_{23}}.
\en
with
 \be \label{eq:v}
 v=\frac{m_{K^+}^2}{m_u+m_s}=\frac{m_K^2-m_\pi^2}{m_s-m_d},
 \en
characterizing the quark-order parameter $\la \bar q q\ra$ which
spontaneously breaks the chiral symmetry.
The nonresonant $\sigma_{_{\rm NR}}$ term is introduced for the
following reason. Although the nonresonant contributions to
$f_s^{KK}$ and $F_s^{KK}$ are related through the equation of
motion, the resonant ones are different and not related {\it a
priori}. As stressed in \cite{CCSKKK}, to apply the equation of
motion, the form factors should be away from the resonant region. In
the presence of resonances, we thus need to introduce a
nonresonant $\sigma_{_{\rm NR}}$ term which can be constrained by
the measured $\overline B^0\to K_SK_SK_S$ rate and the $K^+K^-$ mass
spectrum measured in $\ov B^0\to K^+K^-K_S$ \cite{CCS:nonres}.  The
parameter $\alpha$ appearing in the same equation should be close
to the value of $\alpha_{_{\rm NR}}$ given in Eq.
(\ref{eq:alphaNR}). We will use the experimental measurement
$\alpha=(0.14\pm0.02)\,{\rm GeV}^{-2}$ \cite{BaBarKpKmK0}.

It is known that in the narrow width approximation, the three-body
decay rate obeys the factorization relation
 \be \label{eq:fact}
 \Gamma(B\to RP\to P_1P_2P)=\Gamma(B\to RP)\B(R\to P_1P_2),
 \en
with $R$ being a resonance. This means that the amplitudes $A(B\to
RP\to P_1P_2P)$ and $A(B\to RP)$ should have the same expressions
apart from some factors. Hence, using the known results for
quasi-two-body decay amplitude $A(B\to RP)$, one can have a cross
check on the three-body decay amplitude of $B\to RP\to P_1P_2P$. For example, the factorizable amplitude of the scalar $f_0(980)$ contribution to $B^-\to K^+K^-K^-$ derived from Eq. (\ref{eq:AKpKmKm}) is given by
\be \label{eq:f0KinBtoKKK}
 \la K^+K^-K^-|T_p|
B^-\ra_{f_0} &=& {g^{f_0(980)\to K^+K^-}\over
m_{f_0}^2-s_{23}-im_{f_0}\Gamma_{f_0}}
\Bigg\{ -\bar r_\chi^{f_0}\bar f_{f_0}^sF_0^{BK}(m_{f_0}^2)(m_B^2-m_K^2)
\left(a_6^p-{1\over 2}a_8^p\right) \non \\
&+& f_KF_0^{Bf_0^u}(m_K^2)(m_B^2-m_{f_0}^2)\left[a_1\delta^p_u+a_4^p+a_{10}^p-(a_6^p+a_8^p)r_\chi^K\right]
\Bigg\}.
\en
Comparing this
equation with Eq. (A6) of \cite{CCY:SP}, we see that the expression
inside $\{\cdots\}$ is identical to that of $B^-\to f_0(980)K^-$, as it should be. \footnote{There are some sign typos in Eq. (A6) of \cite{CCY:SP} including the one in the amplitude of $B^-\to f_0K^-$. When comparing Eq. (\ref{eq:f0KinBtoKKK}) with Eq. (A1) of \cite{CCYZ}, we see that some terms are missing in Eq. (\ref{eq:f0KinBtoKKK}). This is because one has to consider the convolution with the light-cone distribution amplitude of the $f_0(980)$ in the approach of QCDF. As a consequence, the amplitude for $f_0$ emission does not vanish in QCDF. We will not consider those subtitles in the simple factorization approach adapted here.} In the above equation, $\bar{r}_\chi^{f_0}=2m_{f_0}/m_b(\mu)$.
The superscript $u$ of the form factor $F_0^{B{f_0}^u}$ reminds
us that it is the $u\bar u$ quark content that gets involved in
the $B$ to $f_0$ form factor transition.

We digress for a moment to discuss the wave function of the
$f_0(980)$. What is the quark structure of the light scalar mesons
below or near 1 GeV has been quite controversial. In this work we
shall consider the conventional $q\bar q$ assignment for the
$f_0(980)$. In the naive quark model, the flavor wave functions of
the $f_0(980)$ and $f_0(500)$ (or $\sigma$ meson) read
 \be
 f_0(500)={1\over \sqrt{2}}(u\bar u+d\bar d), \qquad\qquad f_0(980)= s\bar
 s,
 \en
where ideal mixing for $f_0(980)$ and $f_0(500)$ has been assumed. In
this picture, $f_0(980)$ is purely an $s\bar s$ state. However,
there also exist some experimental evidences indicating that
$f_0(980)$ is not purely an $s\bar s$ state. First, the
observation of $\Gamma(J/\psi\to f_0\omega)\approx {1\over
2}\Gamma(J/\psi\to f_0\phi)$ \cite{PDG} clearly indicates the
existence of the non-strange and strange quark content in
$f_0(980)$. Second, the fact that $f_0(980)$ and $a_0(980)$ have
similar widths and that the $f_0(980)$ width is dominated by $\pi\pi$
also suggests the composition of $u\bar u$ and $d\bar d$ pairs in
$f_0(980)$; that is, $f_0(980)\to\pi\pi$ should not be OZI
suppressed relative to $a_0(980)\to\pi\eta$. Therefore, isoscalars
$f_0(500)$ and $f_0(980)$ must have a mixing
\be \label{eq:fsigmaMix}
 |f_0(500)\ra = -|s\bar s\ra\sin\theta+|n\bar n\ra\cos\theta, \qquad |f_0(980)\ra = |s\bar s\ra\cos\theta+|n\bar n\ra\sin\theta,
\en
with $n\bar n\equiv (\bar uu+\bar dd)/\sqrt{2}$. Experimental
implications for the $f_0(980)\!-\!f_0(500)$ mixing angle have been
discussed in detail in \cite{ChengDSP}. Assuming 2-quark bound states for $f_0(980)$ and $f_0(500)$, the observed large rates of $B^-\to f_0(980)K$ and $f_0(980)K^*$ modes can be explained in QCDF with the mixing angle $\theta$ in the vicinity of $20^\circ$ \cite{CCYZ}.  In this work, we shall use $\theta=20^\circ$.

\begin{table}[!]
\caption{Branching fractions (in units of $10^{-6}$) of resonant and
nonresonant (NR) contributions to $B^-\to K^+K^-K^-$, $\ov B^0\to K^+K^-K^0$, $B^-\to K^-K_SK_S$ and $\ov B^0\to K_SK_SK_S$.}
\begin{ruledtabular} \label{tab:KpKmKm}
\begin{tabular}{l l l l}
 $B^-\to K^+K^-K^-$
   \\
 Decay mode~~
   & BaBar \cite{BaBarKKK}
   & Belle \cite{BelleKpKpKm}
   & Theory
   \\
 \hline
$\phi K^-$
   & $4.48\pm0.22^{+0.33}_{-0.24}$
   &$4.72\pm0.45\pm0.35^{+0.39}_{-0.22}$
   & $2.9^{+0.0+0.5+0.0}_{-0.0-0.5-0.0}$  %
   \\
$f_0(980)K^-$
   & $9.4\pm1.6\pm2.8$
   & $<2.9$
   & $11.0^{+0.0+2.6+0.0}_{-0.0-2.1-0.0}$
   \\
$f_0(1500)K^-$
   & $0.74\pm0.18\pm0.52$
   &
   & $0.62^{+0.0+0.11+0.0}_{-0.0-0.10-0.0}$
   \\
$f_0(1710)K^-$
   & $1.12\pm0.25\pm0.50$
   &
   & $1.1^{+0+0.2+0}_{-0-0.2-0}$
   \\
$f'_2(1525)K^-$
   & $0.69\pm0.16\pm0.13$
   &
   &
   \\
NR
   & $22.8\pm2.7\pm7.6$
   & $24.0\pm1.5\pm1.8^{+1.9}_{-5.7}$
   & $21.8^{+0.8+7.6+0.1}_{-1.1-5.9-0.1}$ %
   \\
\hline
Total
   & $33.4\pm0.5\pm0.9$
   & $30.6\pm1.2\pm2.3$
   &  $26.9^{+0.4+7.5+0.1}_{-0.5-6.1-0.1}$%
   \\
\hline \hline
 $\ov B^0\to K^+K^-\ov K^0$
   \\
Decay mode~~
  & BaBar \cite{BaBarKKK}
  & Belle \cite{Belle2004}
  & Theory
  \\
  \hline
$\phi \ov K^0$
  & $3.48\pm0.28^{+0.21}_{-0.14}$
  &
  & $2.6^{+0.0+0.4+0.0}_{-0.0-0.4-0.0}$ %
  \\
$f_0(980)\ov K^0$
  & $7.0^{+2.6}_{-1.8}\pm2.4$
  &
  & $9.1^{+0.0+1.7+0.0}_{-0.0-1.4-0.0}$
  \\
$f_0(1500)\ov K^0$
  & $0.57^{+0.25}_{-0.19}\pm0.12$
  &
  & $0.55^{+0.0+0.10+0.0}_{-0.0-0.09-0.0}$ 
  \\
$f_0(1710)\ov K^0$
  & $4.4\pm0.7\pm0.5$
  &
  & $1.0^{+0.0+0.2+0.0}_{-0.0-0.2-0.0}$
  \\
$f'_2(1525)\ov K^0$
  & $0.13^{+0.12}_{-0.08}\pm0.16$
  &
  &
  \\
NR
  & $33\pm5\pm9$
  &
  & $12.0^{+0.4+2.8+0.1}_{-0.5-2.4-0.1}$ %
  \\
\hline
Total \footnotemark[1]
  & $25.4\pm0.9\pm0.8$
  & $28.3\pm3.3\pm4.0$
  &  $18.7^{+0.2+3.5+0.0}_{-0.3-3.1-0.0}$%
  \\
\hline \hline
$B^-\to K^-K_SK_S$
  &
  &
  &
  \\
Decay mode~~
  & BaBar \cite{BaBarKKK}
  & Belle \cite{Belle2004}
  & Theory
  \\
  \hline
$f_0(980) K^-$
  & $14.7\pm2.8\pm1.8$
  &
  & $8.7^{+0.0+2.1+0.0}_{-0.0-1.6-0.0}$%
  \\
$f_0(1500)K^-$
  &  $0.42\pm0.22\pm0.58$
  &
  & $0.59^{+0.00+0.10+0.00}_{-0.00-0.09-0.00}$ %
  \\
$f_0(1710)K^-$
  &  $0.48^{+0.40}_{-0.24}\pm0.11$
  &
  & $1.08^{+0.00+0.18+0.00}_{-0.00-0.17-0.00}$ %
  \\
$f'_2(1525)K^-$
  &  $0.61\pm0.21^{+0.12}_{-0.09}$
  &
  \\
NR
  & $19.8\pm3.7\pm2.5$
  &
  & $11.3^{+0.2+3.7+0.0}_{-0.3-3.0-0.0}$%
  \\
\hline
Total
  & $10.1\pm0.5\pm0.3$
  & $13.4\pm1.9\pm1.5$
  & $15.1^{+0.0+3.7+0.0}_{-0.0-3.2-0.0}$ %
  \\
\hline \hline
$\ov B^0\to K_SK_SK_S$
  \\
Decay mode~~
  & BaBar \cite{BaBarKsKsKs}
  & Belle \cite{Belle2004}
  & Theory
  \\
\hline
$f_0(980)K_S$
  & $2.7^{+1.3}_{-1.2}\pm0.4\pm1.2$
  &
  &  $2.4^{+0.0+0.6+0.0}_{-0.0-0.5-0.0}$%
  \\
$f_0(1500)K_S$
  & $$
  &
  &  $0.15^{+0.00+0.03+0.00}_{-0.00-0.02-0.00}$%
  \\
$f_0(1710)K_S$
  & $0.50^{+0.46}_{-0.24}\pm0.04\pm0.10$
  &
  &  $0.28^{+0.00+0.05+0.00}_{-0.00-0.04-0.00}$ %
  \\
$f_2(2010)K_S$
  & $0.54^{+0.21}_{-0.20}\pm0.03\pm0.52$
  &
  &
  \\
NR
  & $13.3^{+2.2}_{-2.3}\pm0.6\pm2.1$
  &
  & $6.58^{+0.09+2.04+0.01}_{-0.12-1.70-0.01}$%
  \\
\hline
Total
  & $6.19\pm0.48\pm0.15\pm0.12$
  & $4.2^{+1.6}_{-1.3}\pm0.8$
  & $6.19^{+0.01+1.62+0.01}_{-0.02-1.42-0.01}$ %
  \\
\end{tabular}
\end{ruledtabular}
\footnotetext[1]{The LHCb measurement is $\B(\ov B^0\to K^+K^-\ov K^0)=(19.1\pm1.5\pm1.1\pm0.8)\times 10^{-6}$ \cite{LHCb:KKpi}.}
\end{table}

Finally, the matrix elements involving three-kaon creation are given
by~\cite{Cheng:2002qu}
 \be \label{eq:KKKme}
 &&\hspace{-0.5cm}\la \ov K {}^0(p_1) K^+(p_2) K^-(p_3)|(\bar s d)_{V-A}|0\ra\la
 0|(\bar d b)_{V-A}|\ov B {}^0\ra
 \approx  0, \\
 &&\hspace{-0.5cm}\la \ov K {}^0(p_1) K^+(p_2) K^-(p_3)|\bar s\gamma_5
 d|0\ra\la
 0|\bar d\gamma_5 b|\ov B {}^0\ra=v\frac{ f_B m_B^2}{f_\pi m_b}
 \left(1-\frac{s_{13}-m_1^2-m_3^2}{m_B^2-m_K^2}\right)F^{KKK}(m_B^2).
 \non
 \en
Both relations in Eq.
(\ref{eq:KKKme}) are originally derived in the chiral limit
\cite{Cheng:2002qu} and hence the quark masses appearing in Eq.
(\ref{eq:v}) are referred to the scale $\sim$ 1 GeV . The first
relation reflects helicity suppression which is expected to be
even more effective for energetic kaons. For the second relation,
we introduce the form factor $F^{KKK}$ to extrapolate the chiral
result to the physical region. Following \cite{Cheng:2002qu} we
shall take $F^{KKK}(q^2)=1/[1-(q^2/\Lambda^2_\chi)]$ with
$\Lambda_\chi=0.83$~GeV being a chiral symmetry breaking scale.

To proceed with the numerical calculations, we shall assume that
the main scalar meson contributions are
those that have dominant $s\bar s$ content and large coupling to
$K\ov K$. We consider the scalar mesons $f_0(980)$, $f_0(1500)$ and $f_0(1710)$
which are supposed to have the
largest couplings with the $K\ov K$ pair. More specifically, we shall use
$g^{f_0(980)\to K^+K^-}=3.7$~GeV,
$g^{f_0(1500)\to K^+K^-}=0.69$~GeV, $g^{f_0(1710)\to K^+K^-}=1.6$~GeV, $\Gamma_{f_0(980)}=80$~MeV,
$\Gamma_{f_0(1500)}=0.109$~GeV, $\Gamma_{f_0(1710)}=0.135$~GeV, $\bar
f_{f_0(980)}(\mu=m_b/2)\simeq 0.46$~GeV~\cite{Cheng:2005ye},
$\bar f_{f_0(1500)}\simeq 0.30$ GeV and $\bar f_{f_0(1710)}\simeq 0.17$ GeV. As for
the parameter $\sigma_{_{\rm NR}}$ in Eq. (\ref{eq:KKssme}), its magnitude can be determined from the
measured $K_SK_SK_S$ rate, namely, $\B(\ov B^0\to
K_SK_SK_S)=(6.1\pm0.5)\times 10^{-6}$ \cite{HFAG}.
As to the strong phase $\phi_r$ we follow \cite{CCS:nonres} to
take $\phi_\sigma\approx \pi/4$ which
yields $K^+K^-$ mass spectrum in $\ov B^0\to K^+K^-K_S$ consistent with the data
\be \label{eq:sigma}
  \sigma_{_{\rm NR}}= e^{i\pi/4}\left(3.39^{+0.18}_{-0.21}\right)\,{\rm GeV}.
\en

The calculated branching fractions of resonant and nonresonant
contributions to $B^-\to K^+K^-K^-$, $\ov B^0\to K^+K^-K^0$, $B^-\to K^-K_SK_S$ and $\ov B^0\to K_SK_SK_S$ are depicted  in
Table \ref{tab:KpKmKm}. The factorizable amplitudes of the last three modes can be found in Appendix A of \cite{CCS:nonres}. Note that both BaBar and Belle used to see a broad scalar resonance $f_X(1500)$ in $B\to K^+K^+K^-$, $K^+K^-K_S$ and $K^+K^-\pi^+$ decays at energies around 1.5 GeV. However, the nature of $f_X(1500)$ is not clear as it cannot be identified with the well known scaler meson $f_0(1500)$.
Nevertheless, the recent angular-momentum analysis of the above-mentioned three channels by BaBar \cite{BaBarKKK} shows that the $f_X(1500)$ state is not a single scalar resonance, but instead can be described by the sum of the well-established resonances $f_0(1500)$, $f_0(1710)$ and $f'_2(1525)$.

From Table \ref{tab:KpKmKm} it is obvious that the predicted rates for
resonant and nonresonant components are consistent with experiment
within errors. It is known that the calculated $\B(B\to \phi K)$ is smaller than experiment and this rate deficit problem calls for the $1/m_b$ power corrections from penguin annihilation. A unique feature of hadronic $B\to KKK$ decays is that they are predominated by the nonresonant contributions with nonresonant fraction of order 80\%. The nonresonant background due to the current-induced process through $B\to KK$ transition accounts only 5\% of the observed nonresonant contributions as it is suppressed by the parameter $\alpha_{_{\rm NR}}$. This implies that the two-body
matrix element of scalar densities e.g. $\langle K\overline K|\bar ss|0\rangle$ induced from the penguin diagram should have a large nonresonant component.
This is plausible because the decay $B\to KKK$ is dominated by the
$b\to s$ penguin transition. Consequently, it is natural to expect that the nonresonant contribution to this decay is also penguin-dominated.

\section{$B\to K\pi\pi$ decays}
The factorizable penguin-dominated $B^-\to K^-\pi^+\pi^-$
decay amplitude has the expression
\be \label{eq:AKmpippim}
 \la K^- \pi^+ \pi^-|T_p|B^-\ra &=&
 \la \pi^+ \pi^-|(\bar u b)_{V-A}|B^-\ra \la K^-|(\bar s u)_{V-A}|0\ra
 \left[a_1 \delta_{pu}+a^p_4+a_{10}^p-(a^p_6+a^p_8) r_\chi^K\right]
 \non\\
&& + \la K^-|(\bar sb)_{V-A}|B^-\ra \la\pi^+\pi^-|(\bar
uu)_{V-A}|0\ra\left[a_2\delta_{pu}+a_3+a_5+a_7+a_9\right] \non \\
&& + \la K^-|(\bar sb)_{V-A}|B^-\ra \la\pi^+\pi^-|(\bar
dd)_{V-A}|0\ra\left[a_3+a_5-{1\over 2}(a_7+a_9)  \right] \non \\
&& + \la K^-|(\bar sb)_{V-A}|B^-\ra \la\pi^+\pi^-|(\bar
ss)_{V-A}|0\ra\left[a_3+a_4^p+a_5-{1\over 2}(a_7+a_9+a_{10}^9)  \right] \non \\
&& +\la K^-|\bar sb|B^-\ra \la\pi^+\pi^-|\bar ss|0\ra
(-2a_6^p+a_8^p) \non \\
&& +\la\pi^-|(\bar db)_{V-A}|B^-\ra \la K^-\pi^+|(\bar
sd)_{V-A}|0\ra(a_4^p-{1\over 2}a_{10}^p) \non \\
&& +\la\pi^-|\bar db|B^-\ra \la K^-\pi^+|\bar
sd|0\ra(-2a_6^p+a_8^p) \non \\
&& +\la K^-\pi^+\pi^-|(\bar su)_{V-A}|0\ra \la0|(\bar
ub)_{V-A}|B^-\ra(a_1\delta_{pu}+a_4^p+a_{10}^p) \non \\
&& +\la K^-\pi^+\pi^-|\bar s(1+\gamma_5)u|0\ra \la0|\bar
u\gamma_5b|B^-\ra(2a_6^p+2a_8^p).
\en
The factorizable amplitudes for other $\ov B\to \ov K\pi\pi$ modes such as $B^-\to\ov
K^0\pi^-\pi^0$, $\ov B^0\to K^-\pi^+\pi^0$, $\ov K^0\pi^+\pi^-$
and $\ov K^0\pi^0\pi^0$ can be found in Appendix A of \cite{CCS:nonres}. The expression of $A(B^-\to K^-\pi^0\pi^0)$ is given in Eq. (\ref{eq:AKmpi0pi0}). All six
channels have the three-body matrix element $\la \pi \pi|(\bar q
b)_{V-A}|B\ra$ which has the similar expression as Eqs. (\ref{eq:r&omega}) and
(\ref{eq:AHMChPT}).
The three-body matrix elements also receive resonant contributions, for example,
\be
 \la K^-(p_1)\pi^+(p_2)|(\bar sb)_\vma|\ov B^0\ra^R &=& \sum_i
{g^{K_i^*\to K^-\pi^+}\over s_{12}-
m_{K_i^*}^2+im_{K_i^*}\Gamma_{K_i^*}}\sum_{\rm
pol}\vp^*\cdot
(p_1-p_2)\la \ov K^{*0}_i|(\bar sb)_\vma|\ov B^0\ra, \non \\
&-&
{g^{K_{0}^*\to K^-\pi^+}\over s_{12}-
m_{K_{0}^*}^2+im_{K_{0}^*}\Gamma_{K_{0}^*}}\la \ov K^{*0}_{0}|(\bar sb)_\vma|\ov B^0\ra,
 \en
with $K_i^*=K^*(892), K^*(1410),K^*(1680),\cdots$, and $K_{0}^*=K_0^*(1430)$.

For the two-body matrix elements $\la \pi^+K^-|(\bar
sd)_{V-A}|0\ra$, $\la \pi^+\pi^-|(\bar uu)_{V-A}|0\ra$ and $\la
\pi^+\pi^-|\bar ss|0\ra$, we note that
 \be
 \la K^-(p_1)\pi^+(p_2)|(\bar sd)_\vma|0\ra &=& \la \pi^+(p_2)|(\bar
 sd)_\vma|K^+(-p_1)\ra = (p_1-p_2)_\mu F_1^{K\pi}(s_{12}) \non \\
 &+& {m_K^2-m_\pi^2\over
 s_{12}}(p_1+p_2)_\mu\left[-F_1^{K\pi}(s_{12})+F_0^{K\pi}(s_{12})\right],
 \en
where we have taken into account the sign flip arising from
interchanging the operators $s\leftrightarrow d$. The resonant contributions are
\be \label{eq:m.e.pole2}
\la K^-(p_1)\pi^+(p_2)|(\bar sd)_\vma|0\ra^R &=& \sum_i
{g^{K^*_i\to K^-\pi^+}\over
s_{12}-m_{K^*_i}^2+im_{K^*_i}\Gamma_{K^*_i}}\sum_{\rm
pol}\vp^*\cdot
(p_1-p_2)\la K^*_i|(\bar sd)_\vma|0\ra \non \\
&-& \sum_i{g^{{K^*_{0i}}\to K^-\pi^+}\over
s_{12}- m_{K^*_{0i}}^2+im_{K^*_{0i}}\Gamma_{K^*_{0i}}}\la
K^*_{0i}|(\bar sd)_\vma|0\ra,
\en
Hence, form factors $F_1^{K\pi}$ and $(-F_1^{K\pi}+F_0^{K\pi})$  receive the following resonant contributions
\be \label{eq:F1Kpi}
 (F^{K\pi}_{1}(s))^R &=& \sum_i{m_{K_i^*}f_{K_i^*}g^{K_i^*\to K\pi}\over
 m_{K_i^*}^2-s-im_{K_i^*}\Gamma_{K_i^*}}, \non \\
 (-F^{K\pi}_1(s)+F^{K\pi}_0(s))^R &=& \sum_i{m_{K_{0i}^*}f_{K_{0i}^*}g^{K_{0i}^*\to K\pi}\over
 m_{K_{0i}^*}^2-s-im_{K_{0i}^*}\Gamma_{K_{0i}^*}}\,{s_{12}\over m_K^2-m_\pi^2}
 \non\\
 &&-\sum_i{m_{K_i^*}f_{K_i^*}g^{K_i^*\to K\pi}\over
 m_{K_i^*}^2-s-im_{K_i^*}\Gamma_{K_i^*}}{s_{12}\over m^2_{K_i^*}}.
\en
Note that for the scalar meson, the decay constant
$\bar f_S$ is defined in Eq. (\ref{eq:decayc}), while $f_S$ is defined by
$\la S(p)|\bar q_2\gamma_\mu q_1|0\ra=f_S p_\mu$. The two decay constants
are related by equations of motion \cite{CCY:SP}
 \be \label{eq:EOM}
 \mu_Sf_S=\bar f_S, \qquad\quad{\rm with}~~\mu_S={m_S\over
 m_2(\mu)-m_1(\mu)},
 \en
where $m_{2}$ and $m_{1}$ are the running current quark masses. The nonresonant contribution $\la
\pi^+(p_2) \pi^-(p_3)|\bar s s|0\ra^{NR}$ vanishes under the OZI
rule.

Now, the amplitude $\la K^-\pi^+|(\bar s d)_{V-A}|0\ra
\la \pi^-|(\bar d b)_{V-A}|B^-\ra$ in Eq. (\ref{eq:AKmpippim}) has the expression
 \be \label{eq:KpiBpi}
&& \la K^-(p_1)\pi^+(p_2)|(\bar s d)_{V-A}|0\ra
\la \pi^-(p_3)|(\bar d b)_{V-A}|B^-\ra  \non\\
&=&F_1^{B\pi}(s_{12})F_1^{K\pi}(s_{12})\left[s_{23}-s_{13}-{(m_B^2-m_\pi^2)(m_K^2-m_\pi^2)
\over s_{12}}\right] \non
\\ &+&  F_0^{B\pi}(s_{12})F_0^{K\pi}(s_{12}){(m_B^2-m_\pi^2)(m_K^2-m_\pi^2)
\over s_{12}},
\en
with
 \be
 \la K^-(p_1) \pi^+(p_2)|\bar s d|0\ra
 =\sum_{i}\frac{m_{{K^*_0}_i} \bar f_{{K^*_0}_i} g^{{K^*_0}_i\to K^-\pi^+}}{m_{{K^*_0}_i}^2-s_{12}-i
 m_{{K^*_0}_i}\Gamma_{{K^*_0}_i}}+\la K^-(p_1) \pi^+(p_2)|\bar s
 d|0\ra^{NR}.
 \en

\begin{table}[!]
\footnotesize{
\caption{Branching fractions (in units of $10^{-6}$) of resonant and
nonresonant (NR) contributions to $B^-\to K^-\pi^+\pi^-$, $B^-\to K^-\pi^0\pi^0$, $\ov B^0\to\ov K^0\pi^+\pi^-$ and $\ov B^0\to K^-\pi^+\pi^0$.  Note that  the BaBar result for
$K_0^{*0}(1430)\pi^-$  in \cite{BaBar:Kmpippim}, $K_0^{*-}(1430)\pi^+$  in \cite{BaBarK0pippim},
all the BaBar results in \cite{BaBarKppimpi0} and Belle results  in \cite{BelleKppimpi0} are their
absolute ones. We have converted them into the product branching
fractions, namely, $\B(B\to Rh)\times \B(R\to hh)$. }
\begin{ruledtabular}
\begin{tabular}{l l l l} \label{tab:Kpipi}
$B^-\to K^-\pi^+\pi^-$
\\
Decay mode~~
  & BaBar \cite{BaBar:Kmpippim}
  & Belle \cite{Belle:Kmpippim}
  &Theory
\\
\hline
$\overline K^{*0}\pi^-$
  & $7.2\pm0.4\pm0.7^{+0.3}_{-0.5}$
  & $6.45\pm0.43\pm0.48^{+0.25}_{-0.35}$
  &  $2.4^{+0.0+0.6+0.0}_{-0.0-0.5-0.0}$
  \\
$\overline K^{*0}_0(1430)\pi^-$
  & $19.8\pm0.7\pm1.7^{+5.6}_{-0.9}\pm3.2$
  &$32.0\pm1.0\pm2.4^{+1.1}_{-1.9}$
  & $11.3^{+0.0+3.3+0.1}_{-0.0-2.8-0.1}$ 
  \\
$\rho^0K^-$ & $3.56\pm0.45\pm0.43^{+0.38}_{-0.15}$
  &$3.89\pm0.47\pm0.29^{+0.32}_{-0.29}$
  & $0.65^{+0.00+0.69+0.01}_{-0.00-0.19-0.01}$
  \\
$f_0(980)K^-$
  & $10.3\pm0.5\pm1.3^{+1.5}_{-0.4}$
  & $8.78\pm0.82\pm0.65^{+0.55}_{-1.64}$
  & $6.6^{+0.0+1.6+0.0}_{-0.0-1.3-0.0}$
  \\
NR
  & $9.3\pm1.0\pm1.2^{+6.7}_{-0.4}\pm1.2$ \footnotemark[1]
  & $16.9\pm1.3\pm1.3^{+1.1}_{-0.9}$
  & $15.5^{+0.0+8.0+0.0}_{-0.0-5.1-0.0}$
  \\
\hline
Total
  & $54.4\pm1.1\pm4.6$
  & $48.8\pm1.1\pm3.6$
  & $33.1^{+0.2+14.3+0.0}_{-0.2-9.2-0.0}$ 
  \\
 \hline \hline
$ B^-\to K^-\pi^0\pi^0$
  \\
Decay mode~~
  & BaBar \cite{BaBarKmpi0pi0}
  &  Belle
  & Theory
  \\
\hline
$K^{*-}\pi^0$
  & $2.7\pm0.5\pm0.4$
  &
  &  $0.91^{+0.00+0.18+0.03}_{-0.00-0.17-0.03}$
  \\
$K^{*-}_0(1430)\pi^0$
  & $$
  &
  & $2.4^{+0.0+0.8+0.0}_{-0.0-0.7-0.0}$
  \\
$f_0(980)K^-$
  & $2.8\pm0.6\pm0.5$
  &
  & $3.3^{+0.0+0.8+0.0}_{-0.0-0.6-0.0}$ 
  \\
NR
  &
  &
  & $5.9^{+0.0+2.5+0.0}_{-0.0-1.8-0.0}$ 
  \\
\hline
Total
  & $16.2\pm1.2\pm1.5$
  &
  &  $11.7^{+0.1+4.2+0.0}_{-0.0-3.1-0.0}$ 
  \\
\hline\hline
$\ov B^0\to \ov K^0\pi^+\pi^-$
  \\
Decay mode~~
  & BaBar \cite{BaBarK0pippim}
  &  Belle \cite{BelleK0pipi}
  & Theory
  \\
\hline
$K^{*-}\pi^+$
  & $5.52^{+0.61}_{-0.54}\pm0.35\pm0.41$
  & $5.6\pm0.7\pm0.5^{+0.4}_{-0.3}$
  &  $2.0^{+0.0+0.5+0.1}_{-0.0-0.5-0.1}$ 
  \\
$K^{*-}_0(1430)\pi^+$
  & $18.5^{+1.4}_{-1.1}\pm1.0\pm0.4\pm2.0$
  & $30.8\pm2.4\pm2.4^{+0.8}_{-3.0}$
  & $10.3^{+0.0+2.9+0.0}_{-0.0-2.5-0.0}$ 
  \\
$\rho^0\ov K^0$
  & $4.37^{+0.70}_{-0.61}\pm0.29\pm0.12$
  & $6.1\pm1.0\pm0.5^{+1.0}_{-1.1}$
  & $0.12^{+0.00+0.49+0.00}_{-0.00-0.07-0.00}$ 
  \\
$f_0(980)\ov K^0$
  & $6.92\pm0.77\pm0.46\pm0.32$
  & $7.6\pm1.7\pm0.7^{+0.5}_{-0.7}$
  & $5.9^{+0.0+1.5+0.0}_{-0.0-1.5-0.0}$
  \\
$f_2(1270)\ov K^0$
  & $1.15^{+0.42}_{-0.35}\pm0.11\pm0.35$
  &
  \\
NR
  & $11.1^{+2.5}_{-1.0}\pm0.9$
  & $19.9\pm2.5\pm1.6^{+0.7}_{-1.2}$
  & $15.0^{+0.2+7.8+0.0}_{-0.2-5.1-0.0}$ 
  \\
\hline
Total
  & $50.2\pm1.5\pm1.8$
  & $47.5\pm2.4\pm3.7$
  & $30.6^{+0.1+13.7+0.0}_{-0.1-8.9-0.0}$ 
  \\
\hline\hline
$\ov B^0\to K^-\pi^+\pi^0$
  \\
Decay mode~~
  & BaBar \cite{BaBarKppimpi0}
  & Belle \cite{BelleKppimpi0}
  & Theory
  \\
\hline
$K^{*-}\pi^+$
  & $2.7\pm0.4\pm0.3$
  & $4.9^{+1.5+0.5+0.8}_{-1.5-0.3-0.3}$
  & $1.0^{+0.0+0.3+0.0}_{-0.0-0.2-0.0}$ 
  \\
$\ov K^{*0}\pi^0$
  & $2.2\pm0.3\pm0.3$
  & $<2.3$
  & $0.7^{+0.0+0.2+0.0}_{-0.0-0.2-0.0}$ 
  \\
$K^{*-}_0(1430)\pi^+$
  & $8.6\pm0.8\pm1.0$
  & $$\footnotemark[2]
  & $5.0^{+0.0+1.5+0.1}_{-0.0-1.2-0.1}$ 
  \\
$\ov K^{*0}_0(1430)\pi^0$
  & $4.3\pm0.3\pm0.7$
  & $$\footnotemark[2] &
  $4.1^{+0.0+1.4+0.0}_{-0.0-1.2-0.0}$ 
  \\
$\rho^+K^-$
  & $6.6\pm0.5\pm0.8$
  & $15.1^{+3.4+1.4+2.0}_{-3.3-1.5-2.1}$
  & $2.4^{+0.0+2.6+0.1}_{-0.0-1.1-0.1}$ 
  \\
NR
  & $7.6\pm0.5\pm1.0$ \footnotemark[3]
  & $5.7^{+2.7+0.5}_{-2.5-0.4}<9.4$
  & $9.0^{+0.3+5.8+0.0}_{-0.3-3.3-0.0}$ 
\\ \hline
Total
  & $38.5\pm1.0\pm3.9$
  & $36.6^{+4.2}_{-4.1}\pm3.0$
  & $18.6^{+0.4+11.9+0.1}_{-0.4-~6.7-0.1}$ 
  \\
\end{tabular}
\end{ruledtabular}
 \footnotetext[1]{The branching fraction  $(2.4\pm0.5^{+1.3}_{-1.5})\times 10^{-6}$ given in Table II of \cite{BaBar:Kmpippim} is for the phase-space nonresonant contribution to $B^-\to K^-\pi^+\pi^-$.}
 \footnotetext[2]{What Belle has measured is for $K_x^*\pi$ where $K_x^*$ is not specified though it could be $K_0^*(1430)$ \cite{BelleKppimpi0}.}
 \footnotetext[3]{The branching fraction $(2.8\pm0.5\pm0.4)\times 10^{-6}$ given in Table VI of \cite{BaBarKppimpi0} is for the phase-space nonresonant contribution to $\ov B^0\to K^-\pi^+\pi^0$.}
 }
\end{table}

We consider the
factorizable amplitude  of the weak decay $B^-\to K_0^{*0}(1430)\pi^-$ followed by the strong decay
$ K^{*0}_0(1430)\to K^-\pi^+$ as a cross
check on the three-body decay amplitude of $B\to RP\to P_1P_2P$.
From Eq. (\ref{eq:AKmpippim}) we obtain
 \be
&& \la K^-(p_1)\pi^+(p_2)\pi^-(p_3)|T_p|
B^-\ra_{K_0^{*0}(1430)} =
\\ && -{g^{K_0^{*0}(1430)\to K^-\pi^+}\over
m_{K_0^*}^2-s_{12}-im_{K_0^*}\Gamma_{K_0^*}}\Bigg\{ \left(
 a_4^p-r_\chi^{K^*_0}a_6^p
 -{1\over 2}(a_{10}^p-r_\chi^{K^*_0}a_8^p) \right)
 f_{K_0^*}F_0^{B\pi}(m_{K_0^*}^2)(m_B^2-m_\pi^2)
 \Bigg\}, \non
 \en
where
 \be
  r^{K^*_0}_\chi(\mu)={2m_{K_0^*}^2\over
 m_b(\mu)(m_s(\mu)-m_q(\mu))}.
 \en
The expression inside $\{\cdots\}$ agrees with the amplitude of $\ov
B^0\to \ov K_0^{*0}(1430)\pi^0$ given in Eq. (A6) of \cite{CCY:SP}.

The momentum dependence of the
weak form factor $F^{K\pi}(q^2)$ is parameterized as
 \be \label{Kpi}
 F^{K\pi}(q^2)=\,{F^{K\pi}(0)\over 1-q^2/{\Lambda_\chi}^2+i\Gamma_R/{\Lambda_\chi}},
 \en
with $\Gamma_R$ being the width of the
relevant resonance, which is taken to be 200 MeV
\cite{Cheng:2002qu}.

\begin{table}[t]
\caption{Branching fractions (in units of $10^{-6}$) of resonant and
nonresonant (NR) contributions to $B^-\to
\ov K^0\pi^-\pi^0$ and $\ov B^0\to\ov K^0\pi^0\pi^0$.}
\begin{ruledtabular} \label{tab:K0pimpi0}
\begin{tabular}{l l | l l }
Decay mode~~~~~~~~~~~~~ & Theory~~~~~~~~~~~~~~~~~~~~~  & Decay mode & Theory \\ \hline
$B^-\to\ov K^0\pi^-\pi^0$
  \\
  \hline
$K^{*-}\pi^0$
  & $1.7^{+0.0+0.3+0.2}_{-0.0-0.3-0.2}$
& $\ov K^{*0}\pi^-$
  & $1.2^{+0.0+0.3+0.0}_{-0.0-0.3-0.0}$
  \\
$K^{*-}_0(1430)\pi^0$
  & $5.4^{+0.0+1.6+0.1}_{-0.0-1.4-0.1}$
& $\ov K^{*0}_0(1430)\pi^-$
  & $5.3^{+0.0+1.6+0.0}_{-0.0-1.4-0.0}$
  \\
$\rho^-\ov K^0$
  & $1.5^{+0.0+2.5+0.0}_{-0.0-0.9-0.0}$
& NR
  & $9.4^{+0.3+6.2+0.0}_{-0.3-3.6-0.0}$ 
  \\
\hline
Total
  & $16.6^{+0.2+10.3+0.0}_{-0.2-~5.8-0.0}$ 
  &
  &
  \\
\hline \hline
$\ov B^0\to \ov K^0\pi^0\pi^0$
  \\
\hline
$f_0(980)\ov K^0$
  & $3.0^{+0.0+0.7+0.0}_{-0.0-0.6-0.0}$
& $\ov K^{*0}\pi^0$
  & $0.88^{+0.00+0.18+0.00}_{-0.00-0.16-0.00}$
  \\
$\ov K^{*0}_0(1430)\pi^0$
  & $2.3^{+0.0+0.8+0.0}_{-0.0-0.6-0.0}$
& NR &
  $5.5^{+0.0+2.3+0.0}_{-0.0-1.7-0.0}$ 
  \\
\hline
Total
  & $10.8^{+0.1+3.9+0.0}_{-0.0-2.9-0.0}$
  \\
\end{tabular}
\end{ruledtabular}
\end{table}

It should be stressed that the nonresonant branching fraction $(2.4\pm0.5^{+1.3}_{-1.5})\times 10^{-6}$ in $B^-\to K^-\pi^+\pi^-$ reported by BaBar \cite{BaBar:Kmpippim} is much smaller than the one $(16.9\pm1.3^{+1.7}_{-1.6})\times 10^{-6}$ measured by Belle (see Table \ref{tab:Kpipi}). Since the BaBar and Belle definitions of the $K_0^*(1430)$ and nonresonant contribution differ, it does not make sense to compare the branching fractions and phases directly.
While Belle
(see e.g. \cite{Belle:Kmpippim}) employed an exponential parametrization to describe the nonresonant contribution, BaBar \cite{BaBar:Kmpippim} used
the LASS parametrization to describe the $K\pi$ $S$-wave and the
nonresonant component by a single amplitude suggested by the LASS
collaboration.  While this
approach is experimentally motivated, the use of the LASS
parametrization is limited to the elastic region of $M(K\pi)\lsim
2.0$ GeV, and an additional amplitude is still required for a
satisfactory description of the data.  In short,
the BaBar definition for the $K^*_0(1430)$ includes an effective range term to account for the low-energy
$K\pi$ S-wave, while for the Belle parameterization, this component is absorbed into the nonresonant piece. For the example at hand, the aforementioned BaBar result ${\cal B}(B^-\to K^-\pi^+\pi^-)_{_{\rm NR}}$  is solely due to the phase-space nonresonant piece.
It is clear that part of the LASS shape is really nonresonant which
has a substantial mixing with $K^*_0(1430)$. In principle, this should be added to the phase-space nonresonant piece to get the total nonresonant contribution.  Indeed, by combining coherently the nonresonant part of the LASS parametrization and the phase-space nonresonant, BaBar found the total nonresonant branching fraction to be  $(9.3\pm1.0\pm1.2^{+6.8}_{-1.3})\times 10^{-6}$.
We see from  Table \ref{tab:Kpipi} that the BaBar result is now
consistent with Belle within errors, though the agreement is not perfect. Likewise, the branching fraction $(2.8\pm0.5\pm0.4)\times 10^{-6}$ of phase-space nonresonant contribution to $\ov B^0\to K^-\pi^+\pi^0$ measured by BaBar \cite{BaBarKppimpi0} is now modified to $(7.6\pm0.5\pm1.0)\times 10^{-6}$ when the nonresonant part of the LASS parametrization is added coherently to the phase-space nonresonant piece (see Table \ref{tab:Kpipi}).

For the resonant contributions from $K_0^*(1430)$,
the branching fractions of the quasi two-body decays $B\to K_0^*(1430)\pi$ can be inferred from Table \ref{tab:Kpipi} and the results are shown in Table \ref{tab:BR2body} below. From the table we see that the measured branching fractions of $K_0^{*-}(1430)\pi^+$ and $K_0^{*0}(1430)\pi^-$ channels are of order $30\times 10^{-6}$ by BaBar and $50\times 10^{-6}$ by Belle.
Note that the BaBar results are obtained from $(K\pi)_0^{*0}\pi^-$ and $(K\pi)_0^{*-}\pi^+$ by subtracting the elastic range term from the $K\pi$ $S$-wave \cite{BaBar:Kmpippim,BaBarK0pippim}. For example, the BaBar result shown in Table \ref{tab:Kpipi} for the branching fraction of
$\overline K^{*0}_0(1430)\pi^-$ comes only from the Breit-Wigner component of the LASS parametrization, while the nonresonant contribution includes both the nonresonant part of the LASS shape and the phase-space nonresonant piece.
Nevertheless, the discrepancy between BaBar and Belle for the $K_0^*\pi$ modes still remains and it is crucial to resolve this important issue.

Experimentally, the nonresonant rates in $B^-\to K^-\pi^+\pi^-$ and $\ov B^0\to \ov K^0\pi^+\pi^-$ are of the same order of magnitude as that in $B\to KKK$ decays (see Tables \ref{tab:KpKmKm} and \ref{tab:Kpipi}). Indeed, this is what we will expect. The nonresonant components of $B\to KKK$ are governed by the $K\ov K$ matrix element  $\la K\ov K|\bar ss|0\ra$. By the same token, the nonresonant contribution to the penguin-dominated $B\to K\pi\pi$ decays should be also dominated by the $K\pi$ matrix element, namely, $\la K\pi|\bar sq|0\ra$. Its precise expression will be given in Eq. (\ref{eq:KpimeNew1}) below.
The reason why the nonresonant fraction is as large as 90\% in
$KKK$ decays, but becomes only $(17\sim 40)\%$ in $K\pi\pi$
channels (see Table \ref{tab:BRexpt}) can be explained as follows.
The nonresonant rates in the $K^-\pi^+\pi^-$ and $\ov
K^0\pi^+\pi^-$ modes should be similar to that in $K^+K^-\ov K^0$
or $K^+K^-K^-$. Since the $KKK$ channel receives resonant
contributions only from $\phi$ and $f_{0}$ mesons, while $K^*,
K^*_{0},\rho,f_{0}$ resonances contribute to $K\pi\pi$ modes,
this explains why the nonresonant fraction is of order 90\% in the
former and becomes of order 40\% or smaller in the latter.

The results of our calculation are shown in Tables
\ref{tab:Kpipi} and VII. It is obvious that except for
$f_0(980)K$,  the predicted rates for $K^{*}\pi$,
$K^{*}_0(1430)\pi$ and $\rho K$ are smaller than the data. Indeed,
the predictions based on QCD factorization for these decays are
also generally smaller than experiment by a factor of 2$\sim$5.
This will be discussed in more details in Sec. VI. As a result, this also explain why our predictions of the total branching fractions of $B\to K\pi\pi$ are smaller than experiment.

\section{$B\to KK\pi$ decays}
In this section we turn to the three-body decay modes $KK\pi$ dominated by $b\to u$ tree and $b\to d$ penguin transitions.
\subsection{$B^-\to K^+K^-\pi^-$ decay}
The factorizable tree-dominated $B^-\to K^+K^-\pi^-$
decay amplitude reads
\be \label{eq:AKpKmpim}
 \la \pi^- K^+ K^-|T_p|B^-\ra &=&
 \la K^+ K^-|(\bar u b)_{V-A}|B^-\ra \la \pi^-|(\bar d u)_{V-A}|0\ra
 \left[a_1 \delta_{pu}+a^p_4+a_{10}^p-(a^p_6+a^p_8) r_\chi^\pi\right]
 \non\\
 &&+\la \pi^-|(\bar d b)_{V-A}|B ^-\ra
                   \la K^+ K^-|(\bar u u)_{V-A}|0\ra
    (a_2\delta_{pu}+a_3+a_5+a_7+a_9)
                   \non\\
 &&+\la \pi^-|(\bar d b)_{V-A}|B ^-\ra
                   \la K^+ K^-|(\bar d d)_{V-A}|0\ra
    \bigg[a_3+a_4^p+a_5-\frac{1}{2}(a_7+a_9+a_{10}^p)\bigg]
    \non\\
 &&+\la \pi^-|(\bar d b)_{V-A}|B ^-\ra
                   \la K^+ K^-|(\bar s s)_{V-A}|0\ra
    \bigg[a_3+a_5-\frac{1}{2}(a_7+a_9)\bigg]
    \non\\
  &&+\la \pi^-|\bar d b|B ^-\ra
                   \la K^+ K^-|\bar dd|0\ra
    (-2a_6^p+a_8^p)
                   \non\\
&&+\la K^-|(\bar s b)_{V-A}|B^-\ra
       \la K^+\pi^-|(\bar d s)_{V-A}|0\ra
       (a^p_4-{1\over 2}a^p_{10})
       \non\\
 &&+\la K^-|\bar s b|B^-\ra
       \la K^+\pi^-|\bar d s|0\ra
       (-2 a^p_6+a^p_8)
       \non\\
  &&  +\la K^+ K^-\pi^-|(\bar d u)_{V-A}|0\ra
     \la 0|(\bar u b)_{V-A}|B^-\ra
       \bigg(a_1\delta_{pu}+a^p_4+a^p_{10}\bigg)
       \non\\
 &&  + \la K^+ K^-\pi^-|\bar d(1+\gamma_5) u|0\ra
       \la 0|\bar u\gamma_5 b|B^-\ra
       (2a^p_6+2a^p_8).
\en
Just as the $B^-\to\pi^-\pi^+\pi^-$ decay, the branching fraction of the nonresonant contribution due to the $b\to u$ tree transition will be too large, of order $42\times 10^{-6}$, if it is evaluated solely based on HMChPT. Hence, the momentum dependence of nonresonant amplitudes in an exponential
form given by Eq. (\ref{eq:ADalitz}) has to be introduced.

Note that we have included the matrix element $\la K^+K^-|\bar
dd|0\ra$. Although its nonresonant contribution vanishes as $K^+$
and $K^-$ do not contain the valence $d$ or $\bar d$ quark, this
matrix element does receive a nonresonant contribution from the scalar $f_0$
pole
 \be
 \la K^+(p_2)K^-(p_3)|\bar dd|0\ra^R
 =\sum_{i}\frac{m_{{f_0}_i} \bar f_{{f_0}_i}^d g^{{f_0}_i\to \pi^+\pi^-}}{m_{{f_0}_i}^2-s_{23}-i
  m_{{f_0}_i}\Gamma_{{f_0}_i}},
 \en
where $\la f_0|\bar dd|0\ra=m_{f_0}\bar f_{f_0}^d$. In the 2-quark
model for $f_0(980)$, $\bar f_{f_0(980)}^d=\bar
f_{f_0(980)}\sin\theta/\sqrt{2}$. Also note that the matrix
element $\la K^-(p_3)|(\bar s b)_{V-A}|B^-\ra \la
\pi^-(p_1)K^+(p_2)|(\bar d
 s)_{V-A}|0\ra$ has a similar expression as Eq. (\ref{eq:KpiBpi})
 \be
&& \la K^-(p_3)|(\bar s b)_{V-A}|B^-\ra \la
\pi^-(p_1)K^+(p_2)|(\bar d
 s)_{V-A}|0\ra \non\\
&=&
-F_1^{BK}(s_{12})F_1^{K\pi}(s_{12})\left[s_{13}-s_{23}-{(m_B^2-m_K^2)(m_K^2-m_\pi^2)
\over s_{12}}\right] \non
\\ && -F_0^{BK}(s_{12})F_0^{K\pi}(s_{12}){(m_B^2-m_K^2)(m_K^2-m_\pi^2)
\over s_{12}}.
 \en
As in Eq. (\ref{eq:F1Kpi}), the form factor $F_1^{K\pi}$ receives
a resonant contribution for the $K^*$ pole.
The nonresonant and various resonant contributions to $B^-\to
K^+K^-\pi^-$ are shown in Table \ref{tab:KKpi}. The predicted
total rate agrees well with experiment.

Note that no clear $\phi(1020)$ signature is observed in the mass region $m_{K^+K^-}^2$ around 1 GeV$^2$ \cite{LHCb:pippippim}. Indeed, the branching fraction of  the  two-body decay $B^-\to\phi\pi^-$
is expected to be very small, of order $4.3\times 10^{-8}$. It is induced mainly from $B^-\to\omega\pi^-$ followed by a small $\omega-\phi$ mixing \cite{CC:Bud}.

\begin{table}[t]
\caption{Predicted branching fractions (in units of $10^{-6}$) of resonant and
nonresonant (NR) contributions to $B^-\to K^+K^-\pi^-$ and  $\ov B^0\to
K_SK^\pm\pi^\mp$. Experimental results are taken from Table \ref{tab:BRexpt}.}
\begin{ruledtabular} \label{tab:KKpi}
\begin{tabular}{l l |l l }
 Decay mode &  &Decay mode &  \\ \hline
$B^-\to K^+K^-\pi^-$
  \\
\hline
$K^{*0}K^-$
  & $0.22^{+0.00+0.04+0.01}_{-0.00-0.04-0.01}$~~~~~~~~ 
& $K^{*0}_0(1430)K^-$
  & $1.0^{+0.0+0.2+0.0}_{-0.0-0.2-0.0}$
  \\
$f_0(980)\pi^-$
  & $0.23^{+0.00+0.01+0.01}_{-0.00-0.01-0.01}$ 
& NR
  & $2.9^{+0.7+0.7+0.0}_{-0.7-0.4-0.0}$ 
  \\
\hline
Total(theory)
  & $5.1^{+0.7+1.1+0.0}_{-0.8-0.7-0.0}$ 
& Total(expt)
  & $5.0\pm0.7$
  \\
\hline \hline
$\ov B^0\to \stackrel{(-)}{K^0} K^\mp\pi^\pm$
  \\
\hline
$K^{*0}\ov K^0$
  & $0.20^{+0.00+0.04+0.00}_{-0.00-0.03-0.00}$
& $K_0^{*0}(1430)\ov K^0$
  & $1.3^{+0.0+0.4+0.0}_{-0.0-0.3-0.0}$
  \\
NR
  & $4.2^{+0.7+1.9+0.1}_{-0.8-0.9-0.1}$  
  \\
\hline
Total(theory)
  & $6.2^{+0.7+2.6+0.1}_{-0.8-1.5-0.1}$ 
& Total(expt)
  & $6.4\pm0.8$
  \\
\end{tabular}
\end{ruledtabular}
\end{table}

\subsection{$\ov B^0\to K_S K^\pm\pi^\mp$ decay}
The factorizable $\ov B^0\to \stackrel{(-)}{K^0} K^\mp\pi^\pm$
decay amplitude is given in Eq. (\ref{eq:AKpK0pim}).
The calculated branching fraction $(6.2^{+2.7}_{-1.7})\times 10^{-6}$ is in good agreement with the current average of BaBar \cite{BaBarKpK0pim} and LHCb \cite{LHCb:KKpi}, namely, $(6.4\pm0.8)\times 10^{-6}$.  The resonant states $K^{*-}$ and $K_0^{*-}(1430)$ are absent in this decay because the quasi two-body decays $\ov B^0\to K^\pm K^{*\mp}$ and $K^\pm K_0^{*\mp}(1430)$ can proceed only through the $W$-exchange diagram and hence they are very suppressed.

\subsection{$\ov B^0\to K^+K^-\pi^0$ decay}
The factorizable amplitude of $\ov B^0\to K^+K^-\pi^0$ can be found in Eq. (\ref{eq:AKpKmpi0}).
Since $\B(B^-\to K^+K^-\pi^-)=(5.0\pm0.7)\times 10^{-6}$ \cite{BaBarKpKmpim}, it has been conjectured that the branching fraction of $\ov B^0\to K^+K^-\pi^0$  should be of order $2.5\times 10^{-6}$, which is indeed very close to the Belle measurement  $(2.17\pm0.65)\times 10^{-6}$ \cite{Belle:KpKmpi0}. However, a detailed study indicates that $\B(\ov B^0\to K^+K^-\pi^0)$ is very small, of order $5\times 10^{-8}$. This is mainly because the short-distance contribution to this mode is much smaller than the $K^+K^-\pi^-$ one because the latter is governed by the external pion-emission tree amplitude, while the former is dominated by the internal pion emission. As a result, $A(\ov B^0\to K^+K^-\pi^0)/A(B^-\to K^+K^-\pi^-)\approx a_2/(\sqrt{2}a_1)$. The experimental observation of a sizable rate for $K^+K^-\pi^0$ implies that this mode should receive dominant long-distance contributions.
Since the branching fraction of $\ov B^0\to \pi^+\pi^-\pi^0$ is of order $20\times 10^{-6}$ (see Table \ref{tab:pippimpi0}), it is tempting to consider a final state rescattering of $\pi^+\pi^-$ into $K^+K^-$ that may substantially enhance the rate of $\ov B^0\to K^+K^-\pi^0$. 
To estimate the effect of $\pi^+\pi^-\to K^+K^-$ rescattering, we work in the framework of \cite{Chua:2007cm} and note that in the quasi-elastic rescattering
in $B\to PP$ modes, the corresponding rescattering amplitude is governed by the so-called annihilation rescatterings. The $K^+K^-$ amplitude receives contributions from the $\pi^+ \pi^-$ amplitude with a rescattering factor of $i(r^{(1/2)}_a+r^{(1/2)}_t)$, where $r_a$ and $r_e$, respectively, correspond to annihilation and total-annihilation rescatterting parameters (see Figs. 1(c), (d),  Eqs. (8) and Eq. (10) of  \cite{Chua:2007cm}). This factor is highly constrained by $\bar B^0\to K^+K^-$ rate and found to be 0.15 in magnitude and $-144^\circ$ in phase \cite{Chua:2007cm}.
Consequently, the contribution to $K^+K^-\pi^0$ rate from $\pi^+\pi^-\pi^0$ rescattering is estimated to be $0.5\times 10^{-6}$, which is too small to account for the observed rate.
Of course, rescattering in three-body is not necessarily the same as the two-body one, but, in general, we do not expect a sizable change from the above estimation. Therefore, the unexpectedly large rate of $\ov B^0\to K^+K^-\pi^0$ still remains unexplained.

\section{Two-body $B\to VP$ and $B\to SP$ decays}

\begin{table}[!]
\caption{Branching fractions (in units of $10^{-6}$) of quasi-two-body decays $B\to VP$ and
$B\to SP$ obtained from the studies of three-body decays based on
the factorization approach. Unless specified, the experimental
results are obtained from the three-body Dalitz plot analyses given in
previous Tables. Theoretical uncertainties have been added in
quadrature. QCD factorization (QCDF) predictions taken from
\cite{CC:Bud} for $VP$ modes and from \cite{CCYZ} for $SP$ channels are
shown here for comparison.}
\begin{ruledtabular} \label{tab:BR2body}
\begin{tabular}{l c c | c c}
 Decay mode~~ &  BaBar & Belle  & QCDF & This work  \\ \hline
$\phi K^-$
  & $9.2\pm0.4^{+0.7}_{-0.5}$
  & $9.6\pm0.9^{+1.1}_{-0.8}$
  & $8.8^{+2.8+4.7}_{-2.7-3.6}$
  & $5.8^{+1.1}_{-1.0}$
  \\
$\phi K^0$
  & $7.1\pm0.6^{+0.4}_{-0.3}$
  & $9.0^{+2.2}_{-1.8}\pm0.7$  \footnotemark[1]
  & $8.1^{+2.6+4.4}_{-2.5-3.3}$
  & $5.3^{+0.9}_{-0.8}$
  \\
$\ov K^{*0}\pi^-$
  & $10.8\pm0.6^{+1.2}_{-1.4}$
  & $9.7\pm0.6^{+0.8}_{-0.9}$
  & $10.4^{+1.3+4.3}_{-1.5-3.9}$
  & $3.6^{+0.9}_{-0.8}$
   \\
$\ov K^{*0}\pi^0$
  & $3.3\pm0.5\pm0.4$
  & $0.4^{+1.9}_{-1.7}\pm0.1$
  & $3.5^{+0.4+1.6}_{-0.4-1.4}$
  & $1.0^{+0.3}_{-0.3}$
  \\
$K^{*-}\pi^+$
  & $8.4\pm0.8$
  & $8.4\pm1.1^{+0.9}_{-0.8}$
  & $9.2^{+1.0+3.7}_{-1.0-3.3}$
  & $3.1^{+0.8}_{-0.7}$
  \\
$K^{*-}\pi^0$
  & $8.2\pm1.5\pm1.1$
  &
  & $6.7^{+0.7+2.4}_{-0.7-2.2}$
  & $2.7^{+0.6}_{-0.5}$
  \\
$K^{*0}K^-$ & $<1.1$
  &
  & $0.80^{+0.20+0.31}_{-0.17-0.38}$
  & $0.33^{+0.06}_{-0.05}$ 
  \\
$\rho^0K^-$
  & $3.56\pm0.45^{+0.57}_{-0.46}$
  & $3.89\pm0.47^{+0.43}_{-0.41}$
  & $3.5^{+2.9+2.9}_{-1.2-1.8}$
  & $0.65^{+0.69}_{-0.19}$
  \\
$\rho^0\ov K^0$ & $4.4\pm0.7\pm0.3$
  & $6.1\pm1.0^{+1.1}_{-1.2}$
  & $5.4^{+3.4+4.3}_{-1.7-2.8}$
  & $0.1^{+0.5}_{-0.1}$
  \\
$\rho^+K^-$ & $6.6\pm0.5\pm0.8$
  & $15.1^{+3.4+2.4}_{-3.3-2.6}$
  & $8.6^{+5.7+7.4}_{-2.8-4.5}$
  & $2.4^{+2.6}_{-1.1}$
  \\
$\rho^-\ov K^0$
  & $8.0^{+1.4}_{-1.3}\pm0.6$ \footnotemark[1]
  &
  & $7.8^{+6.3+7.3}_{-2.9-4.4}$
  & $1.5^{+2.5}_{-0.9}$
  \\
$\rho^0\pi^-$
  & $8.1\pm0.7^{+1.3}_{-1.6}$
  & $8.0^{+2.3}_{-2.0}\pm0.7$  \footnotemark[1]
  & $8.7^{+2.7+1.7}_{-1.3-1.4}$
  & $6.7^{+0.4}_{-0.4}$
  \\
$\rho^\pm\pi^\mp$
  & $22.6\pm1.8\pm2.2$ \footnotemark[1]
  & $22.6\pm1.1\pm4.4$ \footnotemark[1]
  & $25.1^{+1.5+1.4}_{-2.2-1.8}$
  & $17.8^{+3.6}_{-3.2}$
  \\
$\rho^0\pi^0$
  & $1.4\pm0.6\pm0.3$ \footnotemark[1]
  & $3.0\pm0.5\pm0.7$ \footnotemark[1]
  & $1.3^{+1.7+1.2}_{-0.6-0.6}$
  & $1.0^{+0.2}_{-0.1}$
  \\
$f_0(980)K^-;f_0\to \pi^+\pi^-$
  & $10.3\pm0.5^{+2.0}_{-1.4}$ \footnotemark[2]
  & $8.8\pm0.8^{+0.9}_{-1.8}$
  & $8.1^{+1.0+15.4}_{-0.9-~5.5}$ \footnotemark[3]
  & $6.6^{+1.6}_{-1.3}$
  \\
$f_0(980)K^0;f_0\to \pi^+\pi^-$
  & $6.9\pm0.8\pm0.6$
  & $7.6\pm1.7^{+0.8}_{-0.9}$
  & $7.4^{+0.9+14.3}_{-0.8-~5.1}$ \footnotemark[3]
  & $5.9^{+1.5}_{-1.5}$
  \\
$f_0(980)K^-;f_0\to K^+K^-$
  & $9.4\pm1.6\pm2.8$
  & $<2.9$
  &
  & $11.0^{+2.6}_{-2.1}$
  \\
$f_0(980)K^0;f_0\to K^+K^-$
  & $7.0^{+2.6}_{-1.8}\pm2.4$
  &
  &
  & $9.1^{+1.7}_{-1.4}$
  \\
$f_0(980)\pi^-;f_0\to \pi^+\pi^-$
  & $<1.5$
  &
  & $0.13^{+0.02+0.09}_{-0.02-0.06}$ \footnotemark[3]
  & $0.20^{+0.01}_{-0.01}$ %
  \\
$\ov K^{*0}_0(1430)\pi^-$
  & $32.0\pm1.2^{+10.8}_{-~6.0}$
  & $51.6\pm1.7^{+7.0}_{-7.5}$
  & $12.9^{+4.6}_{-3.7}$
  & $18.3^{+8.1}_{-6.5}$
  \\
$\ov K^{*0}_0(1430)\pi^0$
  & $7.0\pm0.5\pm1.1$
  &
  &
 $5.6^{+2.6}_{-1.3}$
  & $6.7^{+3.3}_{-2.7}$
  \\
$K^{*-}_0(1430)\pi^+$
  & $29.9^{+2.3}_{-1.7}\pm3.6$  \footnotemark[4]
  & $49.7\pm3.8^{+6.8}_{-8.2}$
  & $13.8^{+4.5}_{-3.6}$
  & $16.7^{+7.3}_{-5.9}$
  \\
\end{tabular}
\end{ruledtabular}
 \footnotetext[1]{Not determined directly from the Dalitz plot analysis of three-body decays.}
 \footnotetext[2]{The Babar measurement $\B(B^-\to f_0(980)K^-; f_0(980)\to\pi^0\pi^0)=(2.8\pm0.6\pm0.3)\times 10^{-6}$ is not consistent with another BaBar result $\B(B^-\to f_0(980)K^-; f_0(980)\to\pi^+\pi^-)=(10.3\pm0.5^{+2.0}_{-1.4})\times 10^{-6}$ in view of the fact $\B(f_0\to \pi^+\pi^-)=2\B(f_0\to\pi^0\pi^0)$.}
 \footnotetext[3]{We have assumed
$\B(f_0(980)\to\pi^+\pi^-)=0.50$ for the QCDF calculation.}
  \footnotetext[4]{Another BaBar measurement of $\ov B^0\to K^-\pi^+\pi^0$ (see Table \ref{tab:Kpipi}) leads to  $\B(\ov B^0\to K_0^{*-}(1430)\pi^+)=27.8\pm2.5\pm3.3$\,.}
\end{table}

So far we have considered the branching fraction products $\B(B\to
Rh_1)\B(R\to h_2h_3)$ with the resonance $R$ being a vector meson
or a scalar meson. Using the experimental information on $\B(R\to
h_2h_3)$ \cite{PDG}
 \be
&& \B(K^{*0}\to K^+\pi^-)=\B(K^{*+}\to K^0\pi^+)=2\B(K^{*+}\to
K^+\pi^0)={2\over 3}, \non \\
&& \B(K_0^{*0}(1430)\to K^+\pi^-)=2\B(K_0^{*+}(1430)\to
 K^+\pi^0)={2\over 3}(0.93\pm0.10), \non \\ &&\B(\phi\to
 K^+K^-)=0.489\pm0.005\,,
 \en
and applying the narrow width approximation (\ref{eq:fact}),
one can extract the branching fractions of $B\to VP$ and $B\to SP$.
The results are summarized in Table \ref{tab:BR2body}. Except the channels $\rho^-\ov K^0$ from BaBar, $\phi K^0$, $\rho^0\pi^-$ from Belle, $\rho^0\pi^0$ and $\rho^\pm\pi^\mp$ from both BaBar and Belle,
all the experimental results are obtained from the three-body Dalitz plot analyses shown in previous Tables.

We see that except for $\rho\pi$ and $f_0(980)K$ modes, the naive factorization predictions for penguin-dominated decays such as $B\to \phi K, K^*\pi, K_0^*(1430)\pi$ are usually too small by a factor of 2$-$3 and further suppressed for $B\to \rho K$ when confronted with experiment. This calls for $1/m_b$ power corrections to solve the rate deficit problem. Within the framework of QCD factorization, we have considered two different types of power correction effects in order to resolve the \CP puzzles and rate deficit problems with penguin-dominated two-body decays of $B$ mesons  and color-suppressed tree-dominated $\pi^0\pi^0$ and $\rho^0\pi^0$ modes: penguin annihilation and soft corrections to
the color-suppressed tree amplitude \cite{CC:Bud}. However, the consideration of these power corrections for three-body $B$ decays is beyond the scope of this work.

\section{Direct $CP$ asymmetries}

\subsection{Inclusive $CP$ asymmetries}
Experimental measurements of direct \CP violation for various charmless three-body $B$ decays are collected in Table \ref{tab:CPdata}. We notice that
\CP asymmetries of the pair $\pi^-\pi^+\pi^-$ and $K^-K^+K^-$ are of opposite signs, and likewise for the pair $K^-\pi^+\pi^-$ and $\pi^-K^+K^-$. This can be understood in terms of U-spin symmetry.
In the limit of $U$-spin symmetry, $\Delta S$=0 $B^-$ decays can be related to the $\Delta S=1$ one. For example,
\be
A(B^-\to \pi^-\pi^+\pi^-) &=& V_{ub}^*V_{ud}\la \pi^-\pi^+\pi^-|O_d^u| B^-\ra+V_{cb}^*V_{cd}\la \pi^-\pi^+\pi^-|O_d^c|B^-\ra, \non \\
A( B^-\to K^-K^+K^-) &=& V_{ub}^*V_{us}\la K^-K^+K^-|O_s^u| B^-\ra+V_{cb}^*V_{cs}\la K^-K^+K^-|O_s^c|B^-\ra,
\en
where the 4-quark operator $O_s$ is for the $b\to sq_1\bar q_2 $ transition and $O_d$ for the $b\to dq_1\bar q_2$ transition. The assumption of
$U$-spin symmetry implies that under $d \leftrightarrow s$ transitions
\be
\la K^-K^+K^-|O_s^u|B^-\ra=\la \pi^-\pi^+\pi^-|O_d^u|B^-\ra, \qquad
\la K^-K^+K^-|O_s^c|B^-\ra=\la \pi^-\pi^+\pi^-|O_d^c|B^-\ra,
\en
which can be checked from Eqs. (\ref{eq:A3pi}) and  (\ref{eq:AKpKmKm}).
Using the relation for the CKM matrix \cite{Jarlskog}
\be
{\rm Im}(V_{ub}^*V_{ud}V_{cb}V_{cd}^*)=-{\rm Im}(V_{ub}^*V_{us}V_{cb}V_{cs}^*),
\en
it is straightforward to show that
\be
&& |A(B^-\to K^-K^+K^-)|^2-|A(B^+\to K^+K^-K^+)|^2 \non \\
&& =|A(B^-\to \pi^-\pi^+\pi^-)|^2-|A(B^+\to \pi^+\pi^-\pi^+)|^2.
\en
Hence, U-spin symmetry  leads to the relation \cite{Bhattacharya}
\be
R_1\equiv {\A_{CP}(B^-\to \pi^-\pi^+\pi^-)\over  \A_{CP}(B^-\to K^-K^+K^-)} &=& -{\Gamma(B^-\to K^-K^+K^-)\over \Gamma(B^-\to\pi^-\pi^+\pi^-)}.
\en
Likewise,
\be
R_2\equiv {\A_{CP}(B^-\to \pi^-K^+K^-)\over \A_{CP}(B^-\to K^-\pi^+\pi^-)} &=& -{\Gamma(B^-\to K^-\pi^+\pi^-)\over \Gamma(B^-\to\pi^-K^+K^-)}.
\en
The predicted signs of the ratios $R_1$ and $R_2$ are confirmed by experiment.

What is the relative sign between $\A_{CP}(B^-\to \pi^- K^+ K^-)$ and $\A_{CP}(B^-\to \pi^-\pi^+\pi^-)$ ? Applying U-spin symmetry to two of the mesons in final states, one with positive charge and the other with negative charge, we obtain from Eqs. (\ref{eq:A3pi}) and (\ref{eq:AKpKmpim}) that
\be
A(B^-\to \pi^-\pi^-\pi^+)_{p_1p_2p_3}=A(B^-\to \pi^-K^-K^+)_{p_1p_2p_3}+A(B^-\to \pi^-K^-K^+)_{p_2p_1p_3},
\en
where the subscript $p_1p_2p_3$ denotes the momentum of the corresponding meson in order.
Similarly,
\be
A(B^-\to K^-K^-K^+)_{p_1p_2p_3}=A(B^-\to K^-\pi^-\pi^+)_{p_1p_2p_3}+A(B^-\to K^-\pi^-\pi^+)_{p_2p_1p_3}.
\en
The above two relations agree with \cite{Gronau}. Because of the momentum dependence of decay amplitudes, the \CP rate difference in $\pi^-\pi^-\pi^+$ ($K^-K^+K^-$) cannot be related to $\pi^-K^+K^-$ ($K^-\pi^-\pi^+$). Therefore, U-spin or flavor SU(3) symmetry does not lead to any testable relations between $\A_{CP}(\pi^- K^+ K^-)$ and $\A_{CP}(\pi^-\pi^+\pi^-)$ and between $\A_{CP}(K^- \pi^+ \pi^-)$ and $\A_{CP}(K^+K^-K^-)$.

Although symmetry argument alone does not give hints at the relative sign of \CP asymmetries in the pair of $\Delta S=0$ and $\Delta S=1$ decays, a realistic model calculation in the framework of this work shows a positive relative sign. When the unknown two-body matrix elements of scalar densities $\la K\pi|\bar s q|0\ra$ such as $\la K^- \pi^+|\bar sd|0\ra$ and $\la \ov K^0 \pi^-|\bar su|0\ra$, $\la K^- \pi^0|\bar su|0\ra$ and $\la \ov K^0 \pi^0|\bar sd|0\ra$ are related to $\la K^+K^-|\bar ss|0\ra$ via SU(3) symmetry, e.g.
\be \label{eq:Kpim.e.SU3}
 \la K^-(p_1) \pi^+(p_2)|\bar sd|0\ra^{NR}=\la K^+(p_1)K^-(p_2)|\bar
 ss|0\ra^{NR}=f_s^{NR}(s_{12}),
\en
with the expression of $f_s^{NR}$ given in Eq. (\ref{eq:KKssme}), we find
$\A_{CP}(K^-\pi^+\pi^-)\approx -3.7\%$ and $\A_{CP}(K^+K^-\pi^-)\approx 13.1\%$. Hence, they are of the same sign as $\A_{CP}(K^-K^+K^-)$ and $\A_{CP}(\pi^+\pi^-\pi^-)$, respectively. However, the naive predictions are
wrong in signs when confronted with the corresponding data, $(3.3\pm1.0)\%$ and $(-11.9\pm4.1)\%$. That is,
the data in Table \ref{tab:CPdata} indicate that \CP asymmetries of the pair $K^-K^+K^-$ and $K^-\pi^+\pi^-$ are of similar magnitude but opposite in sign and likewise for the pair $\pi^-K^+K^-$ and $\pi^-\pi^+\pi^-$. They have the common feature that when $K^+K^-$ is replaced by $\pi^+\pi^-$, \CP asymmetry is flipped in sign.

Recently, it has been conjectured that maybe the final rescattering between $\pi^+\pi^-$ and $K^+K^-$ in conjunction with {\it CPT} invariance is responsible for the sign change \cite{Bhattacharya,Bigi,Bediaga}. As stressed in \cite{CCSfsi}, the presence of final-state interactions (FSIs) can have an interesting impact on the direct \CP
violation phenomenology. Long-distance final state
rescattering effects, in general, will lead to a different pattern
of \CP violation, namely, ``compound" \CP violation. Predictions
of simple \CP violation are quite distinct from that of compound
\CP violation. Moreover, the sign of \CP
asymmetry can be easily flipped by long-distance rescattering
effects \cite{CCSfsi}. A well known example is the direct \CP violation in $\ov B^0\to K^-\pi^+$. In the heavy quark limit,
the decay amplitudes of charmless two-body decays of $B$ mesons can be described  in terms of decay constants and form factors. However, the predicted direct {\it CP}-violating asymmetries for $\ov B^0\to K^-\pi^+$ and $\ov B_s^0\to K^+\pi^-$ disagree with experiment in signs \cite{BN}. This calls for the the necessity of going beyond the leading $1/m_b$ power expansion. Possible $1/m_b$ power corrections to QCD penguin amplitudes include long-distance charming penguins, final-state interactions and  penguin annihilation. Because of possible ``double counting" problems, one should not take into account all power correction effects simultaneously. It has been shown explicitly in \cite{CCSfsi} that FSIs can account for the sign flip of \CP asymmetry  and the rate deficit of $\ov B^0\to K^-\pi^+$. More precisely, the decays $\ov B^0\to D^{(*)}\ov D_s^{(*)}$ followed by the final-state rescattering $D^{(*)}\ov D_s^{(*)}\to K^-\pi^+$ will give a sizable and negative long-distance contribution $\A_{CP}^{\rm LD}$, so that the net \CP asymmetry $\A_{CP}=\A_{CP}^{\rm SD}+\A_{CP}^{\rm LD}$ is negative for $\ov B^0\to K^-\pi^+$ (for details, see \cite{CCSfsi}).
In the QCD factorization approach \cite{BBNS}, sign flip can be caused by penguin annihilation parameterized in terms of two unknown parameters $\rho_A$ and $\phi_A$.

It is known how to explicitly take into account the constraints from
the {\it CPT} theorem when computing partial rate asymmetries for inclusive decays at the quark level \cite{Hou,Wolf} (for a review, see \cite{Atwood}). However, the implication of the {\it CPT} theorem for \CP asymmetries at the hadron level in exclusive or semi-inclusive reactions is more complicated
and remains mostly unclear \cite{AtwoodCPT}.

Taking the cue from the LHCb observation of $\A_{CP}(\pi^-\pi^+\pi^-)\approx -\A_{CP}(\pi^-K^+K^-)$ and $\A_{CP}(K^-\pi^+\pi^-)\approx -\A_{CP}(K^-K^+K^-)$,
it is conceivable that final-state rescattering may play an important role for direct \CP violation. In the absence of a detailed model of final-state interactions for the pair $B^-\to K^-\pi^+\pi^-$ and $\pi^-K^+K^-$, we shall assume that FSIs amount to giving a large strong phase $\delta$ to the nonresonant component of the matrix element of scalar density  $\la K^-\pi^+|\bar s d|0\ra$
\be \label{eq:Kpime}
 \la K^-(p_1)\pi^+(p_2)|\bar s d|0\ra^{\rm NR}
 = \frac{v}{3}(3 F_{\rm NR}+2F'_{\rm NR})+\sigma_{_{\rm NR}}
 e^{-\alpha s_{12}}e^{i\delta}.
\en
Since \CP violation arises from the interference between tree and penguin amplitudes and since nonresonant penguin contributions to the penguin-dominated decay $K^-\pi^+\pi^-$ are governed by the matrix element $\la K^-\pi^+|\bar sd|0\ra$, it is plausible that a strong phase in  $\la K^-\pi^+|\bar sd|0\ra$ induced from FSIs might flip the sign of \CP asymmetry. A fit to the data of $K^-\pi^+\pi^-$ yields
\be \label{eq:KpimeNew1}
 \la K^-(p_1)\pi^+(p_2)|\bar s d|0\ra^{\rm NR}
 &\approx& \frac{v}{3}(3 F_{\rm NR}+2F'_{\rm NR})+\sigma_{_{\rm NR}}
 e^{-\alpha s_{12}}e^{i\pi}\left(1+4{m_K^2-m_\pi^2\over s_{12}}\right)
\en
with the parameter $\sigma_{_{\rm NR}}$ given in Eq. (\ref{eq:sigma}). It follows from U-spin symmetry that
\be \label{eq:KpimeNew2}
 \la K^+(p_1)\pi^-(p_2)|\bar d s|0\ra^{\rm NR}
 &\approx & \frac{v}{3}(3 F_{\rm NR}+2F'_{\rm NR})+\sigma_{_{\rm NR}}
 e^{-\alpha s_{12}}e^{i\pi}\left(1-4{m_K^2-m_\pi^2\over s_{12}}\right),
\en
which will be used to describe $B\to K\ov K\pi$ decays.
Note that
we have implicitly assumed that power corrections will not affect \CP violation in $\pi^+\pi^-\pi^-$ and $K^+K^-K^-$.

The major uncertainty with direct \CP violation comes from the
strong phases which are needed to induce partial rate \CP
asymmetries. In this work, the strong phases arise from the
effective Wilson coefficients $a_i^p$ listed in Eq. (\ref{eq:ai}), the Breit-Wigner formalism for resonances and the penguin matrix elements of scalar densities. Since direct \CP
violation in charmless two-body $B$ decays can be significantly
affected by final-state rescattering \cite{CCSfsi}, it is natural
to extend the study of final-state rescattering effects to the
case of three-body $B$ decays. We will leave this to a future
investigation.

\begin{table}[t]
\caption{Direct \CP asymmetries (in \%) for various charmless
three-body $B$ decays. Experimental results are taken from \cite{HFAG} and \cite{LHCb:Kppippim,LHCb:pippippim}. The mass regions for local \CP asymmetries are specified in Eqs. (\ref{eq:KKKlocalCP})--(\ref{eq:pipipilocalCP}).}
\begin{ruledtabular} \label{tab:CP}
\begin{tabular}{l  c r }
 Final state~~ & Theory & Experiment \\ \hline
$K^+K^-K^-$
  & $-7.1^{+2.0+1.0+0.1}_{-1.4-1.1-0.1}$ %
  & $-3.7\pm1.0$
  \\
$(K^+K^-K^-)_{\rm region}$
  & $-17.7^{+3.8+2.9+0.3}_{-2.5-3.2-0.3}$ %
  & $-22.6\pm2.2$
  \\
$K^+K^-\pi^-$
  & $-10.0^{+1.5+1.4+0.1}_{-2.4-1.3-0.1}$ 
  & $-12.4\pm4.5$
  \\
$(K^+K^-\pi^-)_{\rm region}$
  & $-18.2^{+0.7+1.7+0.1}_{-1.0-1.5-0.1}$ 
  & $-64.8\pm7.2$
  \\
$K^-\pi^+\pi^-$
  & $2.7^{+0.1+0.7+0.0}_{-0.2-0.8-0.0}$ 
  & $3.3\pm1.0$
  \\
$(K^-\pi^+\pi^-)_{\rm region}$
  & $14.1^{+0.2+13.9+0.4}_{-0.2-11.7-0.4}$
  & $67.8\pm8.5$
  \\
$\pi^+\pi^-\pi^-$
  & $8.7^{+0.5+1.6+0.0}_{-1.1-1.5-0.0}$ %
  & $10.3\pm2.5$
  \\
$(\pi^+\pi^-\pi^-)_{\rm region}$
  & $22.5^{+0.5+2.9+0.1}_{-0.4-3.3-0.1}$ %
  &  $58.4\pm8.7$
  \\
\hline
$K^+K^-K_S$
  & $-5.5^{+1.4+0.5+0.1}_{-1.0-0.5-0.1}$ %
  &
  \\
$K_SK_SK_S$
  & $0.74^{+0.01+0.00+0.01}_{-0.01-0.00-0.01}$ %
  & $17\pm18$
  \\
$K^-K_SK_S$
  & $3.5^{+0.0+0.3+0.1}_{-0.0-0.2-0.1}$ %
  & $4^{+4}_{-5}$
  \\
$K^+K^-\pi^0$
  & $-9.2^{+0.0+0.0+0.0}_{-0.0-0.0-0.0}$
  &
  \\
$K_SK^\pm\pi^\mp$
  & $1.8^{+1.7+1.5+0.0}_{-2.9-2.5-0.0}$ 
  &\\
$\ov K^0\pi^+\pi^-$
  &  $-0.83^{+0.03+0.12+0.01}_{-0.02-0.14-0.01}$
  & $-1\pm5$
  \\
$\ov K^0\pi^-\pi^0$
  &  $0.64^{+0.06+0.04+0.01}_{-0.04-0.06-0.01}$ 
  &
  \\
$\pi^+\pi^-\pi^0$
  & $-1.4^{+0.3+0.5+0.0}_{-0.2-0.7-0.0}$ %
  &
  \\
\end{tabular}
\end{ruledtabular}
\end{table}

The calculated inclusive \CP asymmetries $(8.7^{+1.7}_{-1.9})\%$ for $\pi^+\pi^-\pi^-$ and $(-7.1^{+2.4}_{-1.7})\%$ for $K^+K^-K^-$ (see Table \ref{tab:CP}) are consistent with LHC measurements  in both sign and magnitude (see Table \ref{tab:CPdata}). As noted in passing, if we set $\delta=0$ in Eq. (\ref{eq:Kpime}) so that $\la K^+\pi^-|\bar ds|0\ra=\la K^+K^-|\bar ss|0\ra$, the predicted \CP violation $\A_{CP}(K^-\pi^+\pi^-)=(-3.8^{+1.2}_{-0.7})\%$ will be wrong in sign. If a strong phase $\delta$ is allowed due to some power corrections such as FSIs, we obtain $\A_{CP}(K^-\pi^+\pi^-)=(2.6^{+1.6}_{-1.9})\%$ provided that the modified  matrix element Eq. (\ref{eq:KpimeNew1}) is applied. Using Eq. (\ref{eq:KpimeNew2}) which follows from Eq. (\ref{eq:KpimeNew1}) via U-spin symmetry, we then predict $\A_{CP}(K^+K^-\pi^-)=(-13.4^{+4.6}_{-4.8})\%$ in agreement with experiment.

Besides direct \CP violation in $K^+K^-K^-,K^+K^-\pi^-,K^-\pi^+\pi^-,\pi^-\pi^+\pi^-$ modes, we have calculated {\it CP}-violating asymmetries in other three-body $B$ decays as summarized in Table \ref{tab:CP}. It is expected that $\ov B^0\to K^+K^-\pi^0, K^-\pi^+\pi^0$ and especially $B^-\to\ov K^0\pi^-\pi^0$ can have sizable  asymmetries.

\subsection{Regional $CP$ asymmetries}

Large local \CP asymmetries in three-body charged $B$ decays have been observed by LHCb in the low mass regions specified in Eqs. (\ref{eq:KKKlocalCP})--(\ref{eq:pipipilocalCP}) \cite{LHCb:Kppippim,LHCb:pippippim,deMiranda}. If intermediate resonant states are not associated in these low mass regions, it is natural to expect that the Dalitz plot is governed by nonresonant contributions. In this case direct \CP violation arises solely from the interference of tree and penguin nonresonant amplitudes. For example, in the absence of resonances, \CP asymmetry in $B^-\to K^-\pi^+\pi^-$ stems mainly from the interference of the nonresonant tree amplitude $\la \pi^+\pi^-|(\bar u b)_{V-A}|B^-\ra\la K^-|(\bar su)_{V-A}|0\ra$ and the nonresonant penguin amplitude $\la \pi^-|\bar db|B^-\ra\la K^-\pi^+|\bar sd|0\ra$. The results of the calculated local \CP asymmetries are shown in Table \ref{tab:CPregion}. It is evident that except the mode $K^+K^-\pi^-$, regional \CP violation is indeed dominated by the nonresonant background.

\begin{table}[t]
\caption{Predicted direct \CP asymmetries (in \%) due to nonresonant contributions to  various charmless
three-body charged $B$ decays. The mass regions for local \CP asymmetries are specified in Eqs. (\ref{eq:KKKlocalCP})--(\ref{eq:pipipilocalCP}). LHCb measurements \cite{LHCb:Kppippim,LHCb:pippippim,deMiranda} are shown for comparison. }
\begin{ruledtabular} \label{tab:CPregion}
\begin{tabular}{l  c c c c }
  & $\pi^-\pi^+\pi^-$
  & $K^-\pi^+\pi^-$
  & $K^+K^-\pi^-$
  & $K^+K^-K^-$
  \\ \hline
$(\A_{CP}^{\rm region})_{\rm NR}$
  & $57.4^{+3.2+2.6+1.1}_{-3.4-4.0-1.1}$ %
  & $49.0^{+~7.0+7.7+0.3}_{-10.5-8.4-0.4}$ %
  & $-25.8^{+2.9+2.8+0.4}_{-5.6-2.5-0.4}$ %
  & $-13.2^{+2.0+2.9+0.3}_{-1.2-3.3-0.3}$ %
  \\
$(\A_{CP}^{\rm region})_{\rm expt}$
  & $58.4\pm8.7$
  & $67.8\pm8.5$
  & $-64.8\pm7.2$
  & $-22.6\pm2.2$
\end{tabular}
\end{ruledtabular}
\end{table}

A realistic and straightforward calculation of regional \CP asymmetries in our model yields the results shown in Table \ref{tab:CP}. We see in this table that while regional \CP violation of $K^+K^-K^-$  agrees with experiment within errors,
the predicted local asymmetries of order $-19\%$, 18\% and 23\% for $K^+K^-\pi^-$, $K^-\pi^+\pi^-$ and $\pi^+\pi^-\pi^-$,  respectively, are indeed greatly enhanced with respect to the inclusive ones, though they are still significantly below the corresponding data of order $-65\%$, 68\% and 58\%. The reader may wonder why the realistic calculation yields results different from the naive expectation. We will come to this point later.

It has been claimed recently that the observed large localized \CP violation in $B^-\to \pi^+\pi^-\pi^-$ may result from the interference of a light scalar meson $f_0(500)$  and the vector $\rho^0(770)$ resonance \cite{Zhang,Bhattacharya}, even though the latter resonance is not covered in the low mass region $m^2_{\pi^-\pi^- \rm ~low}<0.4$ GeV$^2$.
Let us first consider the vector meson resonance $\rho^0$ in $B^-\to\pi^+\pi^-\pi^-$ decay. As pointed out in Sec. II.A, the calculated
$\B(B^-\to \rho^0\pi^-)=(6.8\pm0.4)\times 10^{-6}$ is consistent with the world average $(8.3^{+1.2}_{-1.3})\times 10^{-6}$ \cite{HFAG} within errors.
Its \CP asymmetry is found to be $\A_{C\!P}(\rho^0\pi^-)=0.059^{+0.012}_{-0.010}$. At first sight, this seems to be in agreement in sign with the BaBar measurement $0.18\pm0.07^{+0.05}_{-0.15}$ \cite{BaBarpipipi}. However, theoretical predictions based on QCDF, pQCD and soft-collinear effective theory all lead to a negative \CP asymmetry for $B^\mp\to\rho^0\pi^\mp$ (see Table XIII of \cite{CC:Bud}). As shown explicitly in Table IV of \cite{CC:Bud}, within the framework of QCDF, the inclusion of $1/m_b$ power corrections to penguin annihilation is responsible for the sign flip of $\A_{C\!P}(\rho^0\pi^-)$ to a right one. The consideration of power corrections is however beyond the scope of this work based on a simple factorization approach.

As for the scalar resonance $f_0(500)$, if we assume the form factor $F_0^{B\sigma}(0)=0.25$ and take the mixing angle $\theta=20^0$ in Eq. (\ref{eq:fsigmaMix}), we find the branching fraction of $B^-\to f_0(500)\pi^-$ to be order of $2.6\times 10^{-6}$, but its \CP violation is very small, of order $-1\%$. In our model calculation, we find that the local asymmetry due to $\rho^0(770)$ and $f_0(500)$ resonances is $(\A_{CP}^{\rm region})_{\rho+\sigma}\approx -0.02$\,. Of course, the magnitude and even the sign might get modified if the model is improved to yield a negative \CP violation for $B^\mp\to\rho^0\pi^\mp$ as discussed above.

Even the low mass region $m^2_{\pi^-\pi^- \rm ~low}<0.4$ GeV$^2$ is below the resonance $\rho^0(770)$, we find in our calculation $\rho^0(770)$ makes sizable contributions to the rate and \CP violation of $\pi^-\pi^+\pi^-$. Indeed, the fraction of nonresonant  contribution to the total rate is found to be only 10\%.  Therefore, a reliable estimate of \CP violation in the local regions of the Dalitz plot needs to take into account the effects of nearby resonances. As remarked before, our simple factorization model perhaps does not produce the ``right" \CP asymmetry of $B^-\to\rho^0\pi^-$, this may explain why our prediction of $\A_{CP}^{\rm region}$ for $\pi^+\pi^-\pi^-$ is below the LHCb measurement.

For the decay $B^-\to K^+K^-\pi^-$, the resonance $f_0(980)$ is in the low mass region $m^2_{K^+K^-}<1.5$ GeV$^2$, but it is not clear if the intermediate states $K^{*}(892)$ and $K_0^*(1430)$ are excluded. As a result, it is not surprising that the measured (and also the calculated) local asymmetry in this mode is very different from the one arising solely from the nonresonant contribution.

\subsection{Comments on other works}

\CP violation in three-body decays of the charged $B$ meson has been investigated in  Ref. \cite{Bhattacharya,Zhang,He,Bediaga,Zhang:2013iga}. The authors of \cite{Zhang,Bhattacharya}  considered the possibility of having a large local \CP violation in $B^-\to \pi^+\pi^-\pi^-$ resulting from the interference of the resonances $f_0(500)$ and  $\rho^0(770)$. A similar mechanism has been applied to the decay $B^-\to K^-\pi^+\pi^-$ \cite{Zhang:2013iga}.
Studies of flavor SU(3) symmetry imposed on the nonresonant decay amplitudes and its implication on \CP violation were elaborated on in \cite{He}. In our work, we have taken into account both resonant and nonresonant amplitudes simultaneously and worked out their contributions to branching fractions and \CP violation in details. We found that even in the absence of $f_0(500)$ resonance, local \CP asymmetry in $\pi^+\pi^-\pi^-$ can already reach the level of 23\% due to nonresonant and other resonant contributions. Moreover, the regional asymmetry induced solely by the nonresonant component can be as large as 57\% in our calculation.

The strong coupling between $K^+K^-$ and $\pi^+\pi^-$ channels were studied in \cite{Bediaga} to explain the observed asymmetries in $B^-\to K^-K^+K^-$ and $B^-\to K^-\pi^+\pi^-$. Just as the example of $\ov B^0\to K^-\pi^+$ whose \CP violation is originally predicted to have wrong sign in naive factorization and gets a correct sign after power corrections such as final-state interactions or penguin annihilation, are taken into account, it will be very interesting to see  an explicit demonstration of the sign flip of $\A_{CP}(K^-\pi^+\pi^-)$ and $\A_{CP}(\pi^-K^+K^-)$ when the final-state rescattering of $\pi\pi\leftrightarrow K\ov K$ is turned on.

\section{Conclusions}
We have presented in this work a study of charmless three-body decays of $B$ mesons within the framework of  a simple model based on the factorization approach. Our main results are:

\begin{itemize}

\item
Dominant nonresonant contributions to tree-dominated three-body decays arise
from the $b\to u$ tree transition which can be evaluated using heavy meson chiral perturbation theory valid in the soft meson limit. The momentum dependence of nonresonant $b\to u$ transition amplitudes is parameterized in an exponential form
$e^{-\alpha_{_{\rm NR}} p_B\cdot(p_i+p_j)}$ so that the HMChPT
results are recovered in the soft meson limit $p_i,~p_j\to 0$. The parameter $\alpha_{_{\rm NR}}$ is fixed by the measured nonresonant rate in $B^-\to\pi^+\pi^-\pi^-$.

\item A unique feature of hadronic $B\to KKK$ decays is that they are predominated by the nonresonant contributions with nonresonant fraction of order (70-90)\%. It follows  that nonresonant contributions to the penguin-dominated modes should be also dominated by the penguin mechanism. Hence, nonresonant signals must come mainly from the penguin amplitude governed by the matrix element of scalar densities $\la M_1M_2|\bar q_1 q_2|0\ra$.  We use the measurements of $\ov B^0\to K_SK_SK_S$  to constrain the nonresonant component of $\la K\ov K|\bar ss|0\ra$.

\item  The branching fraction of nonresonant contributions is of order $(15-20)\times 10^{-6}$ in penguin-dominated decays $B^-\to K^+K^-K^-,K^-\pi^+\pi^-$ and of order $(3-5)\times 10^{-6}$ in tree-dominated decays $B^-\to \pi^+\pi^-\pi^-, K^+K^-\pi^-$. The nonresonant fraction is predicted to be around 60\% in $B\to K\ov K\pi$ decays.

\item The
intermediate vector meson contributions to three-body decays are
identified through the vector current, while the scalar meson
resonances are mainly associated with the scalar density. Both scalar and vector resonances can contribute to
the three-body matrix element $\la P_1P_2|J_\mu|B\ra$.

\item  The $\pi^+\pi^-\pi^0$ mode is predicted to have a rate
larger than $\pi^+\pi^-\pi^-$ even though the former involves a
$\pi^0$  and has no identical particles in the final state. This is because while the latter is
dominated by the $\rho^0$ pole, the former receives
$\rho^\pm$ and $\rho^0$ resonant contributions.

\item We have made predictions for the resonant and nonresonant contributions to $\ov B^0\to\pi^+\pi^-\pi^0, \ov K^0\pi^0\pi^0, K_SK^\pm\pi^\mp$ and $B^-\to \ov K^0\pi^-\pi^0$.

\item We emphasize that the seemingly huge difference between BaBar and Belle for the nonresonant contributions to $B^-\to K^-\pi^+\pi^-$ and $\ov B^0\to K^-\pi^+\pi^0$ is now relieved when the nonresonant part of the LASS parametrization adapted by BaBar for the description of $K\pi$ $S$-wave is added coherently to the phase-space nonresonant piece.

\item The surprisingly large rate of $\ov B^0\to K^+K^-\pi^0$ observed by Belle is bigger than the naive expectation by two orders of magnitude. It implies that this mode should be dominated by long-distance contributions.
    It may arise from the decay $\ov B^0\to \pi^+\pi^-\pi^0$ followed by the final-state rescattering of $\pi^+\pi^-$ into $K^+K^-$. However, an estimation based on the two-body FSI model shows $\B(\ov B^0\to K^+K^-\pi^0)$ can be enhanced via final-state rescattering only up to the level of $0.5\times 10^{-6}$. Therefore, the unexpectedly large rate of $\ov B^0\to K^+K^-\pi^0$ still remains unexplained.

\item Based on the factorization approach, we have computed the
resonant contributions to three-body decays and determined the rates
for the quasi-two-body decays $B\to VP$ and $B\to SP$. The
predicted $\rho\pi,~f_0(980)K$ and $f_0(980)\pi$ rates are
consistent with experiment, while the calculated $\phi
K,~K^*\pi,~\rho K$ and $K_0^*(1430)\pi$ are too small compared to
the data.

\item While the calculated direct \CP asymmetries for $K^+K^-K^-$ and $\pi^+\pi^-\pi^-$ modes are in good agreement with experiment in both magnitude and sign, the predicted \CP asymmetries in $B^-\to \pi^- K^+K^-$ and $B^-\to K^-\pi^+\pi^-$ are wrong in signs when confronted with experiment. It has been conjectured recently that a possible resolution to this \CP puzzle relies on final-state rescattering of $\pi^+\pi^-$ and $K^+K^-$. Assuming a large strong phase associated with $\la K\pi|\bar sq|0\ra$ arising from some sort of power corrections, we fit it to the data of $K^-\pi^+\pi^-$ and get correct signs for both $\pi^- K^+K^-$ and $K^-\pi^+\pi^-$ modes.  We predict some testable \CP violation in $\ov B^0\to K^+K^-\pi^0$ and  $K^+K^-K_S$.

\item In this work, there are three sources of strong phases: effective Wilson coefficients, propagators of resonances and the matrix element of scalar density $\la M_1M_2|\bar q_1q_2|0\ra$.

\item In the low mass regions devoid of the known resonances, direct \CP violation is naively expected to be dominated by nonresonant contributions. We found that except the $K^+K^-\pi^-$ mode where resonances are not excluded in the local region, partial rate asymmetries due to the nonresonant background are fairly close to the LHCb measurements. However, realistic model calculations show that resonances near the localized region can make sizable contribution to the total rates and asymmetries. At any rate, we have shown that the regional \CP violation is indeed largely enhanced with respect to the inclusive one, though it is still significantly below the data.

\end{itemize}

\vskip 2.0cm \acknowledgments
We are grateful to Xiao-Gang He for useful discussion.
This research was supported in part by the National Center for Theoretical Sciences and the National Science
Council of R.O.C. under Grant Nos. NSC100-2112-M-001-009-MY3 and NSC100-2112-M-033-001-MY3.

\appendix

\section{Input parameters}

Many of the input parameters for the decay constants of  pseudoscalar and vector mesons and form factors for $B\to P,V$ transitions can be found in \cite{CC:Bud} where uncertainties in form factors are shown. The reader is referred to \cite{CCYZ} for decay constants and form factors related to scalar mesons.

For the CKM matrix elements, we use the updated Wolfenstein parameters
$A=0.823$, $\lambda=0.22457$, $\bar \rho=0.1289$ and $\bar
\eta=0.348$ \cite{CKMfitter}. The corresponding CKM angles are
$\sin2\beta=0.689\pm0.019$ and
$\gamma=(69.7^{+1.3}_{-2.8})^\circ$ \cite{CKMfitter}. For the running quark masses we  shall use \cite{PDG,Xing}
 \be \label{eq:quarkmass}
 && m_b(m_b)=4.2\,{\rm GeV}, \qquad~~~~ m_b(2.1\,{\rm GeV})=4.94\,{\rm
 GeV}, \qquad m_b(1\,{\rm GeV})=6.34\,{\rm
 GeV}, \non \\
 && m_c(m_b)=0.91\,{\rm GeV}, \qquad~~~ m_c(2.1\,{\rm GeV})=1.06\,{\rm  GeV},
 \qquad m_c(1\,{\rm GeV})=1.32\,{\rm
 GeV}, \non \\
 && m_s(2.1\,{\rm GeV})=95\,{\rm MeV}, \quad~ m_s(1\,{\rm GeV})=118\,{\rm
 MeV}, \non\\
 && m_d(2.1\,{\rm GeV})=5.0\,{\rm  MeV}, \quad~ m_u(2.1\,{\rm GeV})=2.2\,{\rm
 MeV}.
 \en
Among the quarks, the strange quark gives the major theoretical uncertainty to
the decay amplitude. Hence, we will only consider the uncertainty in the
strange quark mass given by $m_s(2.1\,{\rm GeV})=95\pm5$ MeV.

\section{Decay amplitudes of $B\to PPP$ decays}
Most of the factorizable decay amplitudes of $\Delta S=0$ and $\Delta S=1$ three-body decays $B$ mesons are already collected in Appendix A of \cite{CCS:nonres}. In this work, we have shown the factorizable decay amplitudes of $B^-\to K^+K^-K^-,K^-K^+\pi^-, K^-\pi^+\pi^-,\pi^+\pi^-\pi^-$ for the purpose of discussion and for corrections. In the following we write down the factorizable amplitudes of $B^-\to K^-\pi^0\pi^0$ and $\ov B^0\to K_SK^\pm \pi^\mp,K^+K^-\pi^0$:

\be \label{eq:AKmpi0pi0}
 \la K^- \pi^0 \pi^0|T_p|B^-\ra &=&
 \la \pi^0 \pi^0|(\bar u b)_{V-A}|B^-\ra \la K^-|(\bar s u)_{V-A}|0\ra
 \left[a_1 \delta_{pu}+a^p_4+a_{10}^p-(a^p_6+a^p_8) r_\chi^K\right]
 \non\\
&& + \la K^-\pi^0|(\bar s b)_{V-A}|B^-\ra \la \pi^0|(\bar uu)_{V-A}|0\ra
\left[a_2\delta_{pu}+{3\over 2}(-a_7+a_9)\right] \non \\
&& + \la K^-|(\bar sb)_{V-A}|B^-\ra \la\pi^0\pi^0|(\bar
uu)_{V-A}|0\ra\left[a_2\delta_{pu}+a_3+a_5+a_7+a_9\right] \non \\
&& + \la K^-|(\bar sb)_{V-A}|B^-\ra \la\pi^0\pi^0|(\bar
dd)_{V-A}|0\ra\left[a_3+a_5-{1\over 2}(a_7+a_9)  \right] \non \\
&& + \la K^-|(\bar sb)_{V-A}|B^-\ra \la\pi^0\pi^0|(\bar
ss)_{V-A}|0\ra\left[a_3+a_4^p+a_5-{1\over 2}(a_7+a_9+a_{10}^9)  \right] \non \\
&& +\la K^-|\bar sb|B^-\ra \la\pi^0\pi^0|\bar ss|0\ra
(-2a_6^p+a_8^p) \non \\
&& +\la\pi^0|(\bar ub)_{V-A}|B^-\ra \la K^-\pi^0|(\bar
su)_{V-A}|0\ra(a_4^p+a_{10}^p) \non \\
&& +\la\pi^0|\bar ub|B^-\ra \la K^-\pi^0|\bar
su|0\ra(-2a_6^p-2a_8^p) \non \\
&& +\la K^-\pi^0\pi^0|(\bar su)_{V-A}|0\ra \la0|(\bar
ub)_{V-A}|B^-\ra(a_1\delta_{pu}+a_4^p+a_{10}^p) \non \\
&& +\la K^-\pi^0\pi^0|\bar s(1+\gamma_5)u|0\ra \la0|\bar
u\gamma_5b|B^-\ra(2a_6^p+2a_8^p),
\en
\be \label{eq:AKpK0pim}
 \la \buildrel (-)\over {K^0} K^\mp\pi^\pm  |T_p|\ov B^0\ra &=&
 \la K^+ \ov K^0|(\bar u b)_{V-A}|\ov B^0\ra \la \pi^-|(\bar d u)_{V-A}|0\ra
 \left[a_1 \delta_{pu}+a^p_4+a_{10}^p-(a^p_6+a^p_8) r_\chi^\pi\right]
 \non\\
 &&+\la \pi^+|(\bar u b)_{V-A}|\ov B ^0\ra
                   \la K^- K^0|(\bar d u)_{V-A}|0\ra
    (a_1\delta_{pu}+a_4^p+a_{10}^p)
                   \non\\
   &&+\la K^-\pi^+|(\bar s b)_{V-A}|\ov B^0\ra
       \la K^0|(\bar d s)_{V-A}|0\ra  \left[a^p_4-{1\over 2}a^p_{10}-(a^p_6-{1\over 2}a^p_8) r_\chi^K\right]
       \non\\
&&+\la \ov K^0|(\bar s b)_{V-A}|\ov B^0\ra
       \la K^+\pi^-|(\bar d s)_{V-A}|0\ra
       (a^p_4-{1\over 2}a^p_{10})
       \non\\
 &&+\la \pi^+|\bar u b|\ov B ^0\ra
   \la K^- K^0|\bar d u|0\ra (-2a^p_6-2a^p_8)
                   \non\\
 &&+\la \ov K^0|\bar s b|\ov B^0\ra
       \la K^+\pi^-|\bar d s|0\ra
       (-2 a^p_6+a^p_8)
       \non\\
  &&  +\la \stackrel{(-)}{K^0} K^\mp\pi^\pm|(\bar uu)_{V-A}|0\ra
     \la 0|(\bar d b)_{V-A}|\ov B^0\ra
       \bigg(a_2\delta_{pu}+a_3+a_5+a_7+a_9\bigg)
       \non\\
 &&  + \la \stackrel{(-)}{K^0} K^\mp\pi^\pm|\bar d(1+\gamma_5) d|0\ra
       \la 0|\bar d\gamma_5 b|\ov B^0\ra
       (2a^p_6-a^p_8),
\en

\be \label{eq:AKpKmpi0}
 \la \pi^0 K^+ K^-|T_p|\ov B^0\ra &=&
 \la \pi^0|(\bar d b)_{V-A}|\ov B^0\ra
                   \la K^+ K^-|(\bar u u)_{V-A}|0\ra
    (a_2\delta_{pu}+a_3+a_5+a_7+a_9)
                   \non\\
  &&+\la \pi^0|\bar d b|\ov B^0\ra
                   \la K^+ K^-|\bar dd|0\ra
    (-2a_6^p+a_8^p)  \non\\
 &&+\la \pi^0|(\bar d b)_{V-A}|\ov B ^0\ra
                   \la K^+ K^-|(\bar s s)_{V-A}|0\ra
    \bigg[a_3+a_5-\frac{1}{2}(a_7+a_9)\bigg]
    \non\\
   &&  +\la K^+ K^-\pi^0|(\bar u u)_{V-A}|0\ra
     \la 0|(\bar d b)_{V-A}|\ov B^0\ra
       \bigg(a_2\delta_{pu}+a^p_4+a^p_{10}\bigg)
       \non\\
 &&  + \la K^+ K^-\pi^0|\bar d\gamma_5 d|0\ra
       \la 0|\bar d\gamma_5 b|\ov B^0\ra
       (2a^p_6-a^p_8).
\en


\end{document}